\documentclass[prd, preprint,amsmath,amssymb,showpacs,showkeys,superscriptaddress,nofootinbib, 12pt, twocolumn]{revtex4-2}
\usepackage{times}
\usepackage{amsmath}
\usepackage{graphicx}
\usepackage{epsfig}
\usepackage{dcolumn}
\usepackage{bm}
\usepackage{color}
\usepackage{longtable}
\usepackage{array}
\usepackage{multirow}
\usepackage{setspace}
\usepackage[mathlines]{lineno}
\usepackage{epsfig,graphics,subfigure,psfrag}
\usepackage{makecell}
\usepackage{geometry}
\usepackage{epstopdf}
\usepackage{verbatim}

\usepackage[pdfstartview = FitH, colorlinks=red, urlcolor = blue,citecolor = blue, linkcolor = green]{hyperref}
\usepackage[hyphenbreaks]{breakurl}

\newcommand{\PPJ}{e^{+}e^{-} \rightarrow \pi^{+}\pi^{-}J/\psi}
\newcommand{\EE}{e^{+}e^{-}}
\newcommand{\MM}{\mu^{+}\mu^{-}}
\newcommand{\LL}{\ell^{+}\ell^{-}}
\newcommand{\PP}{\pi^{+}\pi^{-}}

\emergencystretch=\maxdimen
\begin{document}

\title{\boldmath{Study of the resonance structures in $e^{+}e^{-} \rightarrow \pi^{+}\pi^{-}J/\psi$ process}}

\author{
\author{Author list}
		\begin{small}
        \begin{center}
M.~Ablikim$^{1}$, M.~N.~Achasov$^{10,b}$, P.~Adlarson$^{68}$, M.~Albrecht$^{4}$, R.~Aliberti$^{28}$, A.~Amoroso$^{67A,67C}$, M.~R.~An$^{32}$, Q.~An$^{64,50}$, X.~H.~Bai$^{58}$, Y.~Bai$^{49}$
% ...
\\
\vspace{0.2cm}
(BESIII Collaboration)\\
\vspace{0.2cm} {\it
$^{1}$ Institute of High Energy Physics, Beijing 100049, People's Republic of China\\
% ...
$^{4}$ Bochum Ruhr-University, D-44780 Bochum, Germany\\
% ...
$^{10}$ G.I. Budker Institute of Nuclear Physics SB RAS (BINP), Novosibirsk 630090, Russia\\
% ...
$^{28}$ Johannes Gutenberg University of Mainz, Johann-Joachim-Becher-Weg 45, D-55099 Mainz, Germany\\
% ...
$^{32}$ Liaoning Normal University, Dalian 116029, People's Republic of China\\
% ...
$^{49}$ Southeast University, Nanjing 211100, People's Republic of China\\
$^{50}$ State Key Laboratory of Particle Detection and Electronics, Beijing 100049, Hefei 230026, People's Republic of China\\
% ...
$^{58}$ University of Jinan, Jinan 250022, People's Republic of China\\
% ...
$^{64}$ University of Science and Technology of China, Hefei 230026, People's Republic of China\\
% ...
$^{67}$ University of Turin and INFN, (A)University of Turin, I-10125, Turin, Italy; (B)University of Eastern Piedmont, I-15121, Alessandria, Italy; (C)INFN, I-10125, Turin, Italy\\
$^{68}$ Uppsala University, Box 516, SE-75120 Uppsala, Sweden\\
% ...
$^{b}$ Also at the Novosibirsk State University, Novosibirsk, 630090, Russia\\
% ...
}\end{center}
 \vspace{0.4cm}
 \end{small}
 \
}


\begin{abstract}
  \par Using about 23\ $\mathrm{fb^{-1}}$ of data collected with the BESIII detector operating at the BEPCII storage ring, a precise measurement of the $\PPJ$ Born cross section is performed at center-of-mass energies from 3.7730 to 4.7008\ GeV. Two structures, identified as the $Y(4220)$ and the $Y(4320)$ states, are observed in the energy-dependent cross section with a significance larger than $10\sigma$. The masses and widths of the two structures are determined to be ($M,~\Gamma$) = ($4221.4\pm1.5\pm2.0$\ MeV/$c^{2}$,~$41.8\pm2.9\pm2.7$\ MeV) and ($M,~\Gamma$) = ($4298\pm12\pm26$\ MeV/$c^{2}$, $127\pm17\pm10$\ MeV), respectively. A small enhancement around 4.5\ GeV with a significance about $3\sigma$, compatible with the $\psi(4415)$, might also indicate the presence of an additional resonance in the spectrum. The inclusion of this additional contribution in the fit to the cross section affects the resonance parameters of the $Y(4320)$ state.
\end{abstract}

\maketitle


\section{Introduction}
\par A series of charmonium-like states, commonly referred to as $XYZ$ states, has been observed in the last two decades. The $Y(4260)$ was first observed by BaBar using an initial state radiation (ISR) technique in the process $\EE \rightarrow \gamma_{\rm{ISR}} \PP J/\psi$~\cite{BaBar:2005hhc} and was soon confirmed by Belle in the same process~\cite{Belle:2007dxy}. In 2017, the BESIII collaboration was able to resolve the $Y(4260)$ structure into a combination of two resonances, the $Y(4220)$ and the $Y(4320)$, using the world's largest sample of $\PPJ$ events~\cite{BESIII:2016bnd}. Moreover, similar structures have been observed in the processes $\EE \rightarrow \PP \psi(2S)$~\cite{BESIII:2017tqk}, $\PP h_{c}$~\cite{BESIII:2016adj}, $\pi^{+}D^{0}D^{\ast -}$~\cite{BESIII:2018iea}, $\eta J/\psi$~\cite{BESIII:2020bgb} and $\omega \chi_{cJ}$ $(J=0,\ 1,\ 2)$~\cite{BESIII:2014rja, BESIII:2015som} with the BESIII data. The parameters of the two resonances are to a large extent consistent among reactions. However, to understand whether the structures observed in different final states are indeed the same, more investigations are needed.

\par Conventional charmonium states, such as the $\psi(4040)$, $\psi(4160)$, and $\psi(4415)$, mainly decay into open charm final states ($D^{(\ast)}\bar{D}^{(\ast)}$), while the $Y$ states show strong coupling to the hidden-charm final states~\cite{Mo:2006ss}. The number of observed vector states in this energy region exceeds that of the predicted vector charmonium states~\cite{Charmodel}. These features suggest that some of these supernumerary vector states are candidates of an exotic nature, such as hybrid, tetraquark states, or mesonic molecules ~\cite{Berwein:2015vca, Epelbaum:2014sea, Guo:2013zbw, Cleven:2013mka, Ayala:2018ulm, Giron:2020fvd}. To clarify the nature of these states and to distinguish between the different theoretical models, precise measurements of the production cross section and of the resonance parameters are essential.

\par In this paper, an improved measurement of the energy-dependent $\PPJ$ cross section at center-of-mass energies ($\sqrt{s}$) between 3.7730 to 4.7008\ GeV with a total integrated luminosity of about 23\ fb$^{-1}$ is presented. The $XYZ$ data and the $R$-scan data (between $\sqrt{s}=4.410$ and $4.590$~GeV) analyzed in Ref.~\cite{BESIII:2016bnd} are also analyzed here, and thus these measurements are correlated and can not be used in combination. Improvements in this paper include the use of additional data in the $Y(4220)$/$Y(4320)$ mass region, allowing us to study these two states in more detail, as well as the inclusion of more detailed background studies and systematic studies of the event selection.

\section{The BESIII detector and data sets}
The BESIII detector~\cite{bes3} records $\EE$ collisions provided by the BEPCII storage ring~\cite{bepcii} in a center-of-mass energy range from 2.0 to 4.9 GeV. The cylindrical core of the detector covers 93\% of the full solid angle and consists of a helium-based multilayer drift chamber~(MDC), a plastic scintillator time-of-flight system~(TOF), and a CsI(Tl) electromagnetic calorimeter~(EMC), which are all enclosed in a superconducting solenoidal magnet providing a 1.0~T magnetic field. The solenoid is supported by an octagonal flux-return yoke with resistive plate counter muon identification modules interleaved with steel. The charged-particle momentum resolution at $1~{\rm GeV}/c$ is $0.5\%$, and the $\mathrm{d}E/\mathrm{d}x$ resolution is $6\%$ for electrons from Bhabha scattering. The EMC measures photon energies with a resolution of $2.5\%$ ($5\%$) at $1$~GeV in the barrel (end cap) region. The time resolution in the TOF barrel region is 68 ps, while that in the end cap region is 110 ps.  The end cap TOF system was upgraded in 2015 using multi-gap resistive plate chamber technology, providing a time resolution of 60 ps~\cite{etof}.

\par The data samples analyzed in this article are listed in appendix. They include 40 energy points (referred to as the $XYZ$ data sample) from 3.7730 to 4.7008\ GeV with an integrated luminosity of more than 40 $\rm {pb^{-1}}$ at each $\sqrt{s}$, and 13 energy points (the $R$-scan data sample) from 4.410 to 4.590\ GeV with a luminosity of $7-9$ $\rm {pb^{-1}}$ at each $\sqrt{s}$ to study possible structure around 4.5 GeV. The integrated luminosities and $\sqrt{s}$ are measured using Bhabha scattering, radiative di-muon events, and $\Lambda_{c}$ pairs~\cite{BESIII:2022xii, BESIII:2020eyu, Lambdac}.

\par The {\sc geant4}-based~\cite{GEANT4} Monte Carlo (MC) simulation software packages {\sc boost}~\cite{boost} and  {\sc evtgen}~\cite{EvtGen} are used to determine detection efficiencies and to estimate the background contributions. The generator {\sc kkmc}~\cite{Jadach:2000ir} is used to model the beam energy spread and the ISR emission in $\EE$ annihilations. Final state radiation (FSR) from charged (final state) particles is incorporated using the {\sc photos} package~\cite{photos}. The signal MC samples of $\PPJ$, with $J/\psi \rightarrow \LL~(\ell = e,~\mu)$, are generated using a phase-space (PHSP) model and weighted following the results of a partial wave analysis (PWA) of the data. The amplitude model and the PWA framework are the same as in Refs.~\cite{Ablikim:2020pzw, BESIII:2017bua, pwa}. Potential background contaminations are studied via inclusive MC samples described in Refs.~\cite{BESIII:2016bnd, Zhou:2020ksj}. Bhabha, di-muon, and $\EE\rightarrow 2(\PP)$ events are chosen as control samples to study the tracking efficiencies of $e^{\pm}$, $\mu^{\pm}$, and $\pi^{\pm}$, respectively.
An exclusive MC sample for $e^{+}e^{-} \rightarrow e^{+}e^{-}\mu^{+}\mu^{-}$ (with $\gamma^{\ast}\gamma^{\ast} \rightarrow \mu^{+}\mu^{-}$, two-photon process)~\cite{Berends:1986if, Berends:1986ig} is generated for Boosted Decision Tree (BDT) training~\cite{tmva} used for background suppression.

\section{Event selection}
Events with exactly four charged tracks and zero net charge are selected. In order to be considered, a charged track reconstructed in the MDC is required to satisfy $|\rm{cos}\theta|<0.93$, with $\theta$ being the polar angle, and the distance of closest approach to the $e^{+}e^{-}$ interaction point must be within $\pm10$ cm in the beam direction and within 1 cm in the plane perpendicular to the beam direction. If there are only two tracks with momentum greater than 1.0~GeV$/c$, then the two tracks are judged as leptons; if there are three tracks with momentum greater than 1.0~GeV$/c$, then the two tracks with opposite charges and with an invariant mass closer to the $J/\psi$ mass~\cite{PDG} are regarded as leptons. The other two tracks are regarded as pions. The energy deposited in the EMC is used to distinguish between electrons and muons. For muon candidates, the deposited energy is required to be less than 0.4 GeV, while for electrons, it is required to be larger than 1.1 GeV. In order to suppress the background contribution and to improve the energy and momentum resolution, a four-constraint (4C) kinematic fit is applied to the event with the hypothesis $\EE \rightarrow \PP\LL$, which constrains the total four-momentum of the final state particles to that of the initial colliding beams. The $\chi^{2}$ of the kinematic fit is required to be less than 60. The cosine of the opening angle of the pion-pair ($\rm cos(\pi^{+}\pi^{-})$) and, only in the $J/\psi \rightarrow \EE$ decay channel, of the pion-electron pairs ($\rm cos(\pi^{\pm}e^{\mp})$) are required to be less than 0.98 to suppress gamma conversion background events of the radiative Bhabha and di-muon background contributions~\cite{BESIII:2016bnd}.

\par
\begin{figure*}
\centering
  \includegraphics[width=0.98\textwidth, height=0.35\textwidth]{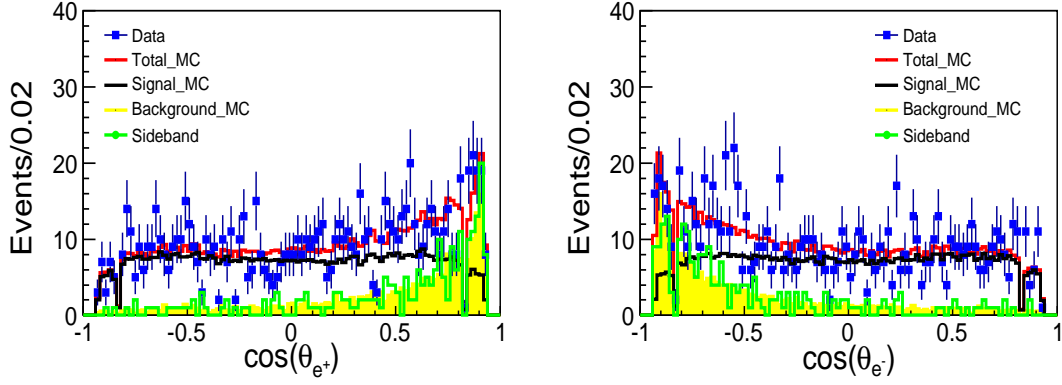}
\caption{\footnotesize{Cosine of the polar angle of the $e^{+}$ ($\rm{cos}(\theta_{e^{+}})$) and $e^{-}$} ($\rm{cos}(\theta_{e^{-}})$) from the sample at  $\sqrt{s} = 4.1780$ GeV. The blue dots with error bars are data from the signal region ($3.08< M(\ell^{+}\ell^{-})<3.12$ GeV/$c^{2}$). The black line is the signal MC, which is normalized to the number of signal events. The green line comes from the sideband region ($3.02<M(\LL)<3.06$ GeV/$c^{2}$ or $3.14<M(\LL)<3.18$ GeV/$c^{2}$ , with an event weight of 0.5) of $M(\LL)$. The yellow-filled area is the background MC (two-photon process MC), and it is normalized to the number of events in the sideband area.  The red line is the sum of signal MC plus background MC simulation.}
\label{Cost}
\end{figure*}

 In order to reduce the background contributions from low momentum electrons, pions are identified using the $\mathrm{d}E/\mathrm{d}x$ information recorded by the MDC. A discriminator $\chi_{\pi^{+/-}} \equiv (\mu_{\rm m}-\mu_{\rm exp})/\sigma_{\rm m}$ is defined by combining the measured $\mathrm{d}E/\mathrm{d}x$ value ($\mu_{\rm m}$), the measurement uncertainty ($\sigma_{\rm m}$), and the expected value under a pion hypothesis ($\mu_{\rm exp}$). The conditions $\chi_{\pi^{+}}<2.8~(3.2)$ and $\chi_{\pi^{-}}<3.0~(4.10)$ for the $\EE$ mode ($\MM$ mode) are found to provide an optimal balance between signal efficiency loss and background rejection power.
The polar angle distributions of $e^{+}$ and $e^{-}$ in the $J/\psi \rightarrow \EE$ mode (Fig.~\ref{Cost}) clearly show a significant contribution from the $e^{+}e^{-} \rightarrow e^{+}e^{-}\mu^{+}\mu^{-}$ process (a two-photon process corresponding to the yellow filled area in Fig.~\ref{Cost}). A boosted decision tree method, implemented within the TMVA framework~\cite{tmva}, is trained to efficiently suppress this background.

\par
\begin{figure*}
\centering
  \includegraphics[width=0.98\textwidth, height=0.4\textwidth]{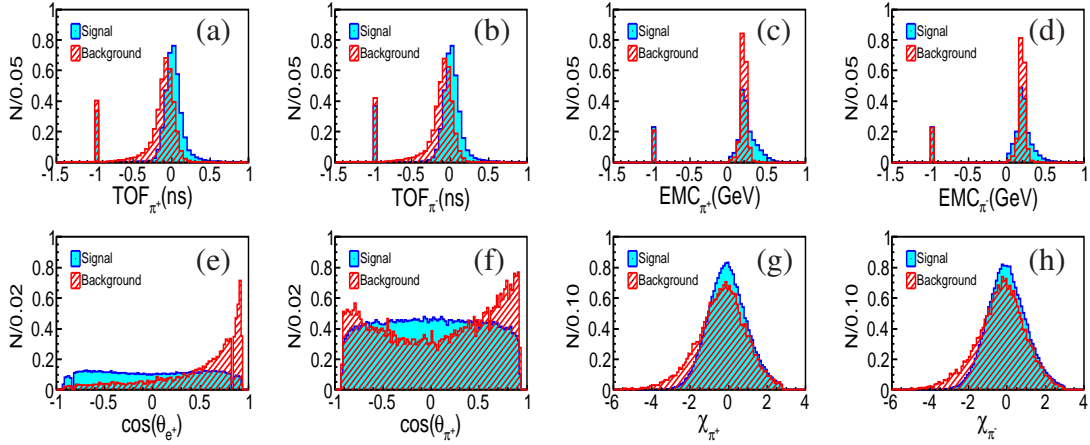}
  \put(-345, 155){(a)}
  \put(-240, 155){(b)}
  \put(-133, 155){(c)}
  \put(-30,  155){(d)}
  \put(-345, 68){(e)}
  \put(-240, 68){(f)}
  \put(-133, 68){(g)}
  \put(-30,  68){(h)}
\caption{\footnotesize{The variables used in the BDT. (a,b) Distribution of the difference between the measured value and the expected value from the TOF of the two pions. (c,d) The energy deposited in the EMC by the two pions. (e,f) The cosine of the polar angle of $e^{+}$ ($\rm{cos}(\theta_{e^{+}})$) and $\pi^{+}$ ($\rm{cos}(\theta_{\pi^{+}})$). (g,h) Distributions of $\chi_{\pi^{+}}$ and $\chi_{\pi^{-}}$. The blue line is for the signal and the red line for the background. In (a,b,c,d), the peak at $-1.0$ corresponds to the tracks without TOF or EMC information. The background and the signal have the same magnitude. All the distributions come from the signal MC and background MC simulation.}}
\label{BDT1}
\end{figure*}

\begin{figure*}
\centering
  \begin{minipage}{1.0\linewidth}
  \includegraphics[width=0.49\textwidth, height=0.45\textwidth]{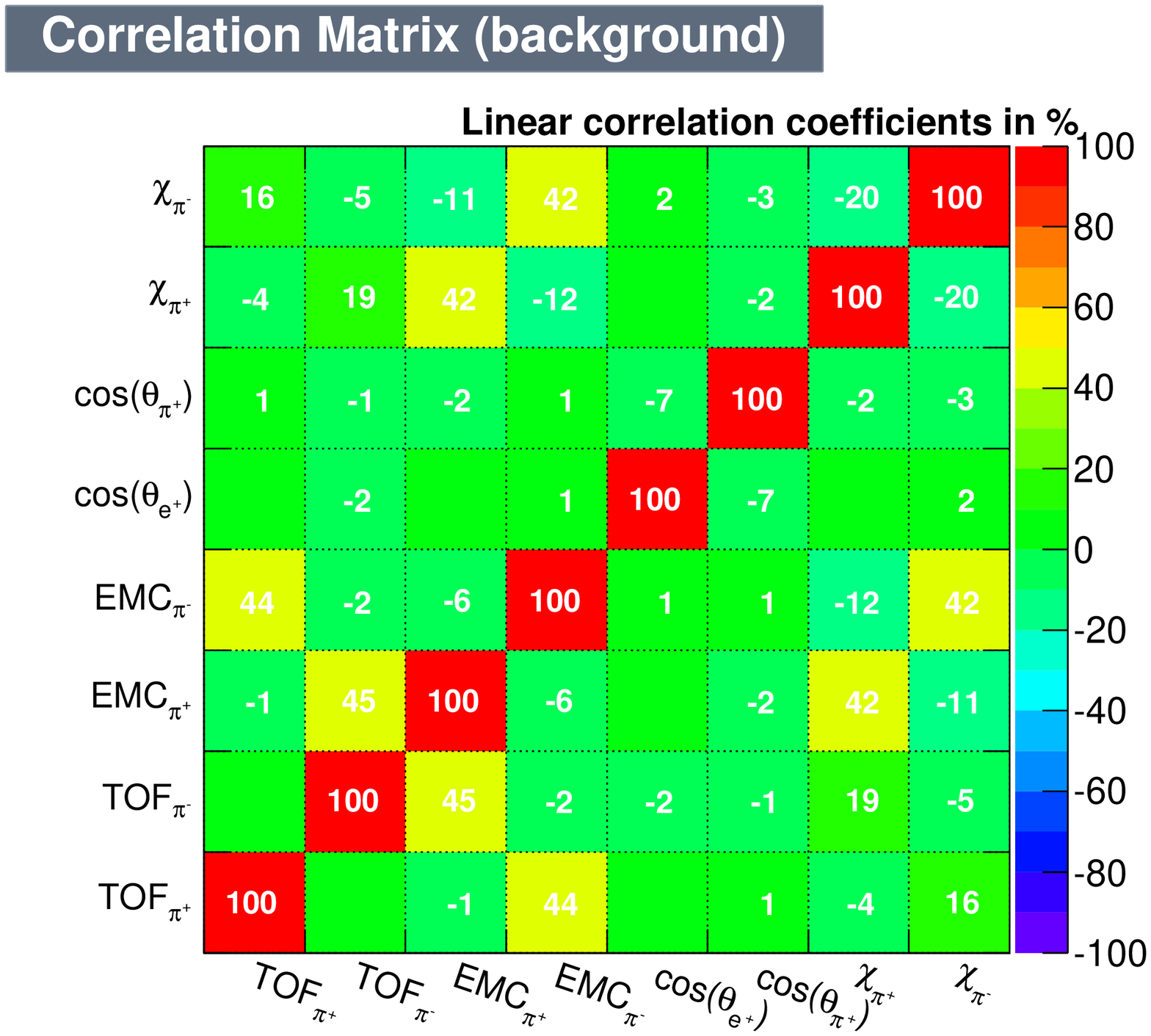}
  \includegraphics[width=0.49\textwidth, height=0.45\textwidth]{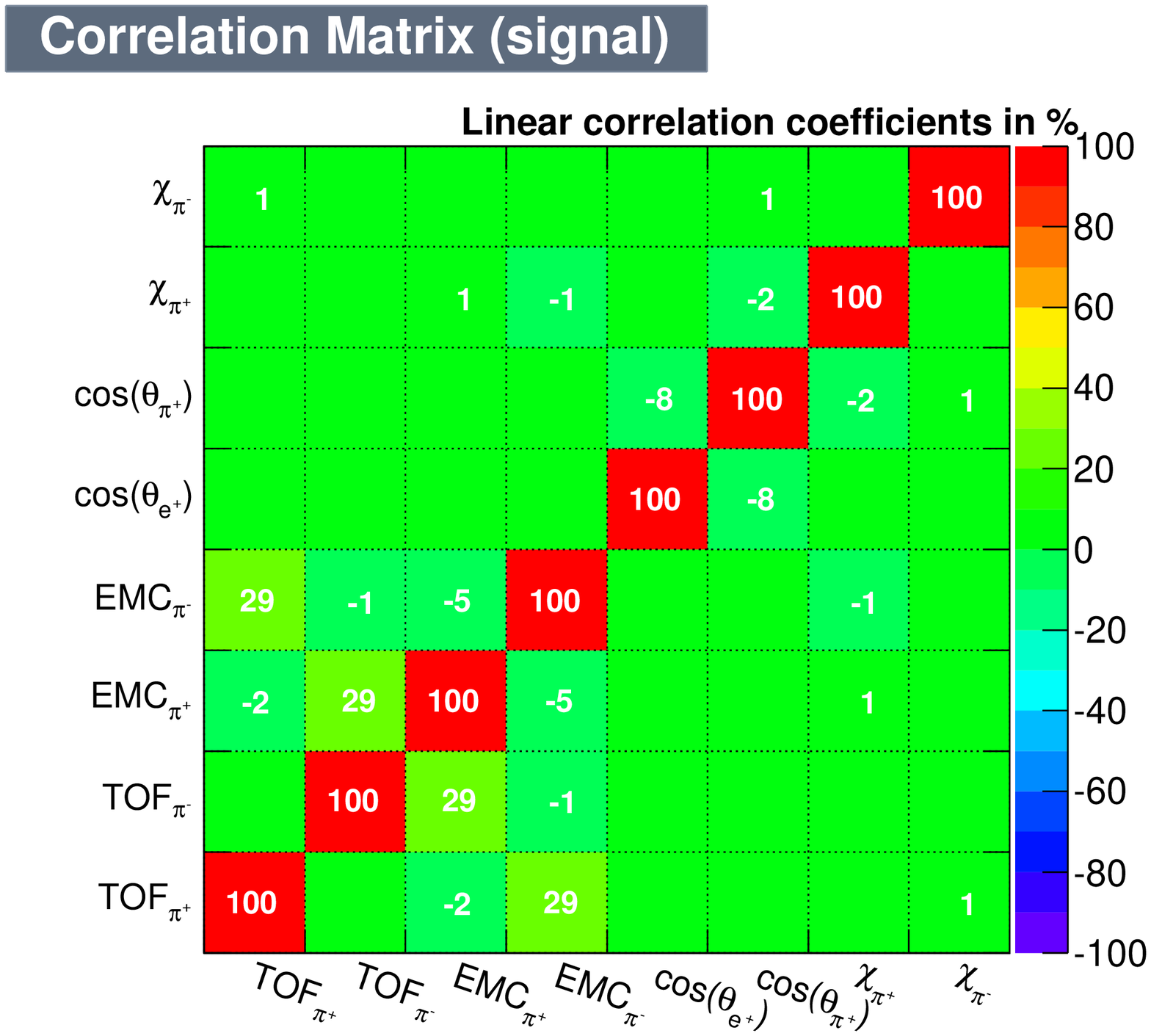}
  \end{minipage}
  \put(-320, -90){(a)}
  \put(-110, -90){(b)}

  \begin{minipage}{1.0\linewidth}
  \includegraphics[width=0.49\textwidth, height=0.45\textwidth]{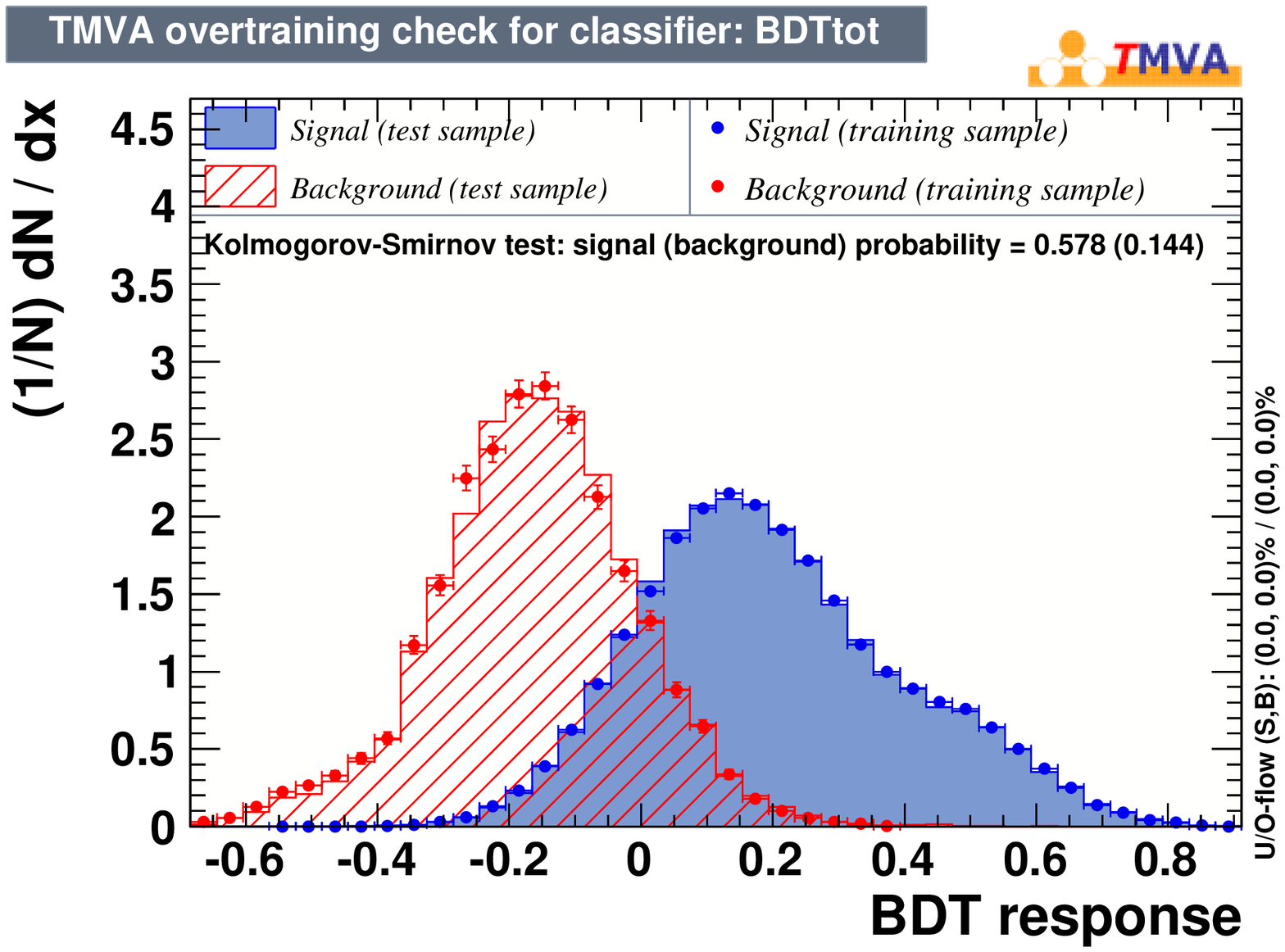}
  \includegraphics[width=0.49\textwidth, height=0.45\textwidth]{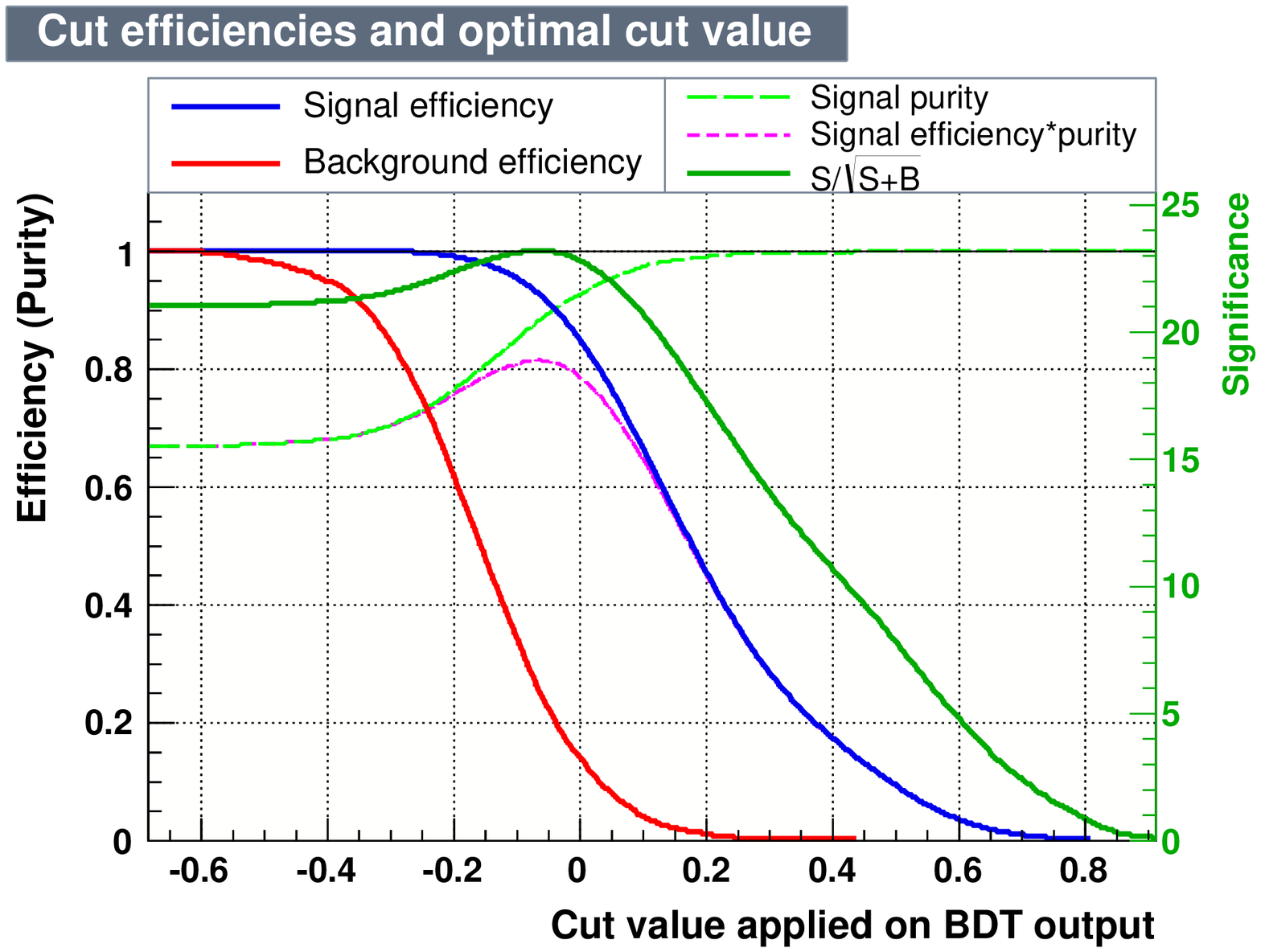}
  \end{minipage}
  \put(-320, -100){(c)}
  \put(-110, -100){(d)}
\caption{\footnotesize{(a,b) The linear correlation coefficients of the variables in the BDT training: (a) is for signal MC and (b) is for background MC simulation samples. (c) The overtraining check of the classifier: the training samples and test samples are in good agreement. (d) The optimized results of the sample at $\sqrt{s} = 4.1780$ GeV. The numbers of signal and background events estimated from the data are used to maximize $S/\sqrt{S+B}$ and thus optimize the event selection criteria.}}
\label{BDT2}
\end{figure*}

 The energy deposited in the EMC, the time of flight from TOF, $\mathrm{d}E/\mathrm{d}x$, and the opening angle of the pion and electron candidates (Fig.~\ref{BDT1}) are used as input variables to the BDT, as they are mostly uncorrelated, as can be seen from Figs.~\ref{BDT2}(a) and (b). The BDT is trained using the MC simulation of the process $e^{+}e^{-} \rightarrow e^{+}e^{-}\mu^{+}\mu^{-}$. The signal MC and background MC simulation of multiple energy points are combined to train the BDT model. Figure~\ref{BDT2}(c) shows that the training samples and test samples in the BDT model are in good agreement, and the model can effectively distinguish between signal and background.
The response of the BDT is required to maximize
the $S/\sqrt{S+B}$ distribution (where the $S$ and $B$ indicate the number of signal and background events passing the condition) shown in Fig.~\ref{BDT2}(d).

\begin{figure*}
\centering
  \begin{minipage}{1.0\linewidth}
  \includegraphics[width=0.98\textwidth, height=0.35\textwidth]{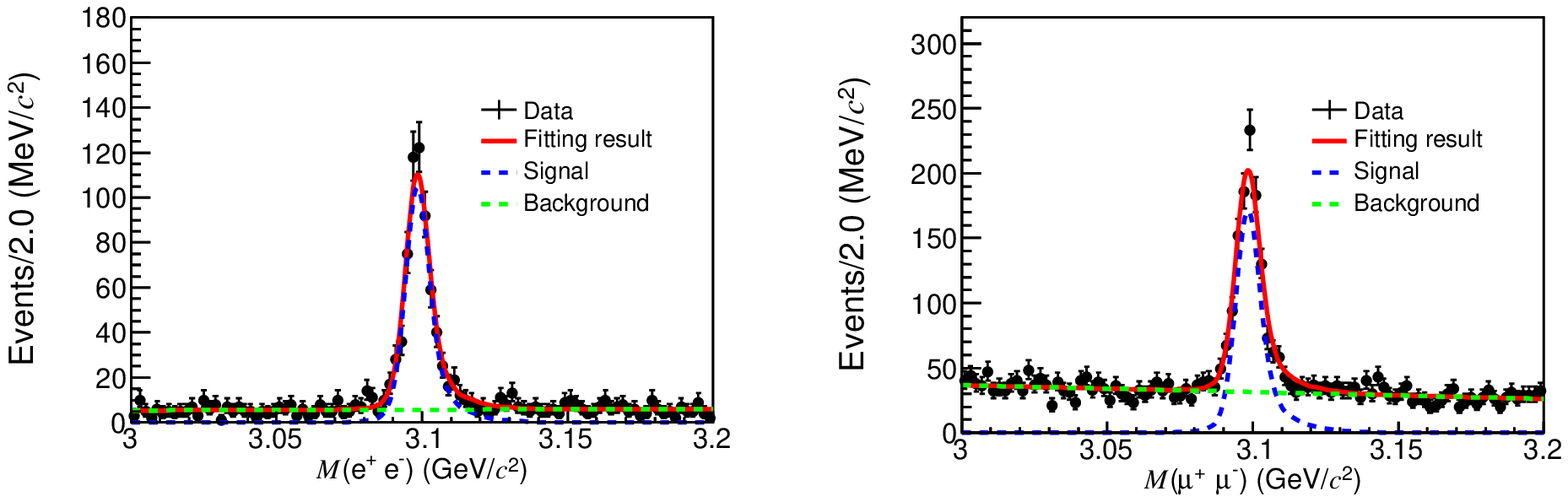}
  \end{minipage}
  \put(-380, 55){(a)}
  \put(-165, 55){(b)}
  \put(-363, 53){\small{$\sqrt{s} = 4.1780$ GeV}}
  \put(-148, 53){\small{$\sqrt{s} = 4.1780$ GeV}}

  \begin{minipage}{1.0\linewidth}
   \includegraphics[width=0.98\textwidth, height=0.4\textwidth]{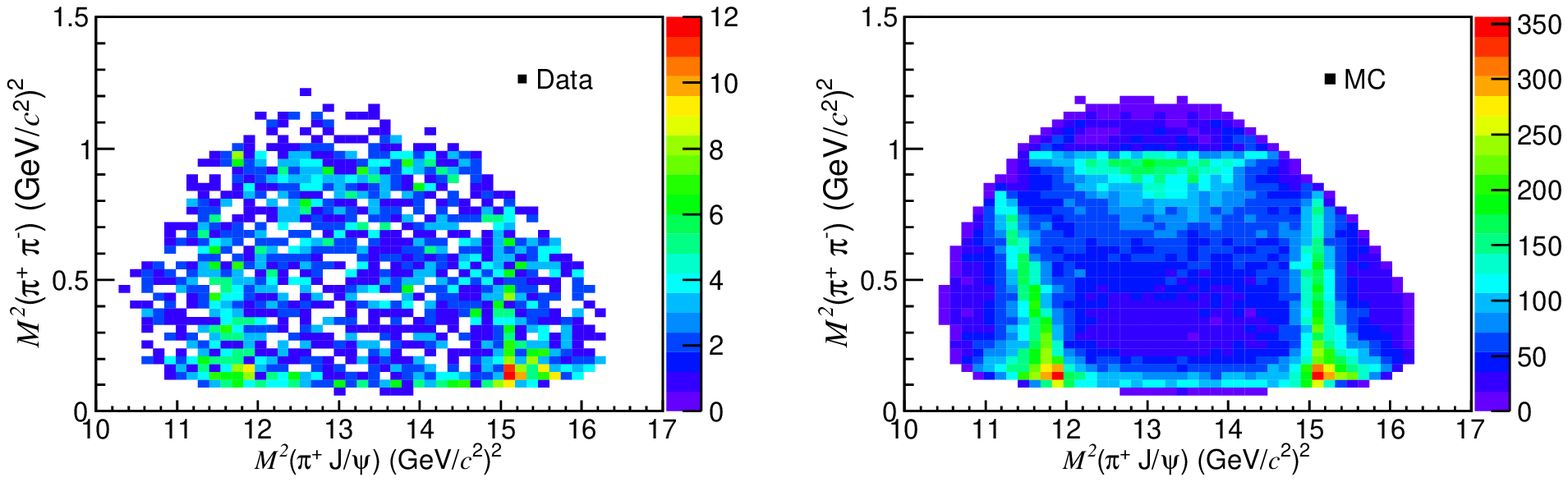}
  \end{minipage}
  \put(-380, 53){(c)}
  \put(-165, 53){(d)}
\caption{\footnotesize{The (a) and (b) plots show the fit results of the invariant mass of the $\EE$ mode and $\MM$ mode, respectively.  The (c) and (d) are the two-dimensional distributions of the squared invariant mass of the $\pi^{+}\pi^{-}$ and $\pi^{+}J/\psi$ pairs. Here, (c) is the data, (d) is the MC simulation from PWA. The contents shown come from the sample at $\sqrt{s} = 4.1780$ GeV.}}
\label{InvariantMass}
\end{figure*}

\section{Measurement of the dressed Born cross section}
\par After applying the event selection criteria mentioned above, a clear $J/\psi$ signal peak is observed in the invariant mass distribution of the lepton pairs ($M(\ell^{+}\ell^{-})$), as shown in Figs.~\ref{InvariantMass}(a)($\EE$ mode) and (b)($\MM$ mode), Figs.~\ref{InvariantMass}(c)(data) and (d)(MC) show the two-dimensional distribution of the invariant mass of the $\pi^{+}\pi^{-}$ and $\pi^{+}J/\psi$ pairs.
To determine the signal yields, an unbinned maximum-likelihood fit to $M(\ell^{+}\ell^{-})$ is performed. The signal probability density function (PDF) is defined as the convolution between the MC signal and a Gaussian function, while the background contribution is parameterized with a linear term.

\par The dressed Born cross section of $\PPJ$ at a given $\sqrt{s}$ is calculated with

\begin{equation}
\label{Cscalculate}
    \sigma(\sqrt{s}) = \frac{N^{\rm obs}}{\mathcal{L}_{\rm int}(1+\delta)\epsilon\mathcal{B}},
\end{equation}
where $N^{\rm obs}$ is the number of signal events, $\mathcal{L}_{\rm int}$ is the integrated luminosity, $\epsilon$ is the selection efficiency, $\mathcal{B}$ is the branching fraction of $J/\psi \rightarrow \ell^{+}\ell^{-}$~\cite{PDG} and ($1+\delta$) is the ISR correction factor. The final cross section is given by the average of the $e^{+}e^{-}$ and the $\mu^{+}\mu^{-}$ modes, weighted with the inverse of the respective statistical uncertainties. The parameters of Eq.~\ref{Cscalculate} and the measured cross sections at the different $\sqrt{s}$ are summarized in the appendix.
\par The ISR correction factor is evaluated using an iterative procedure, in which the $\sigma(\sqrt{s})$ is initially assumed to be simply flat and iteratively recomputed until the difference between the final two iterations is less than 0.1\%. The $(i+1)^{\rm th}$ iteration $(1+\delta)^{i+1}$ is given by~\cite{Sun:2020ehv}:
\begin{equation}
    (1+\delta)^{i+1} = (1+\delta)^{0} \sum^{N}_{j}{W_{j}^{i}}/N,~W_{j}^{i}=\frac{\sigma^{i}(m_{j})}{\sigma^{i}(\sqrt{s})},
\end{equation}
 where $W_{j}^{i}$ is the weighting factor for the $j^{\rm th}$ event. The ISR correction factor $(1+\delta)^{0}$ corresponds to the $0^{\rm th}$ iteration, i.e. to an assumed flat shape of the cross section; $\sigma^{i}$ is the line-shape given by the $i^{\rm th}$ iteration and $\sigma^{0}$ is calculated with the correction factor $(1+\delta)^{0}$, which is calculated using the {\sc kkmc}~\cite{Jadach:2000ir} program; $m_{j}$ is the invariant mass of the final state of the event $j$; and $N$ is the total number of generated MC events. Considering that the sizes of the $R$-scan data samples are small, only the $XYZ$ data samples are used in the iteration. The ISR correction factors for the $R$-scan data samples are calculated from the model describing the cross section in Sect.~\ref{lineshapes}.

\section{Systematic uncertainties of the cross section measurements}
\par The main contributions to the systematic uncertainty of the cross section measurement are related to the measurement of the luminosity, the MC model, the tracking efficiency, the ISR correction, the branching fractions of $J/\psi$ decays, the kinematic fit, the fit to $M(\LL)$, and the BDT method. The integrated luminosities of all data sets are measured using large angle Bhabha scattering events with an uncertainty of 0.66\%~\cite{BESIII:2022xii}. The branching fraction of $J/\psi \rightarrow \ell^{+}\ell^{-}$ is taken from PDG~\cite{PDG}, and the uncertainties are 0.6\% for each mode. The uncertainty related to the kinematic fit is estimated by adjusting the helix parameters of the charged tracks in MC simulation to match the $\chi^{2}$ distribution of data and MC~\cite{BESIII:2012mpj}. The uncertainty is determined as the difference between the results before and after the helix parameters correction, resulting in 0.76\% (1.34\%) for the $\mu^{+}\mu^{-}$ ($e^{+}e^{-}$) mode. The uncertainty on the signal yield arising from the fitting of the $M(\ell^{+}\ell^{-})$ distribution is obtained by varying the fit range and by changing the background modeling from a first to a second order polynomial, which leads to a difference of 2.44\% (1.16\%) that is taken as the systematic uncertainty introduced by the fit method.

\par In the evaluation of the systematic uncertainties due to the tracking efficiency, the MC simulation model, the BDT method, and the ISR correction, 300 sets of Gaussian samplings according to the related central value ($\mu_{0}$) and statistical uncertainties, obtained from the control samples, the PWA fitted parameters, the data-MC simulation ratios of the BDT training variables, and the ISR correction factors, are generated. A Gaussian function ($\mu_{1},~\sigma_{1}$) is used to fit the resulting distributions from the 300 MC simulation samples. The final uncertainty of each source is given by $(|\mu_{1} - \mu_{0}|+\sigma_{1})/\mu_{0}\times100\%$.

\par The uncertainty of the tracking efficiency comes from the uncertainties of the correction factors obtained from control samples, which are the Bhabha, di-muon and $\EE \rightarrow 2(\PP)$ processes. The correction factors are given by the ratio of data to MC simulation in the two dimensions of transverse momentum and the cosine of the polar angle. Therefore, the Gaussian sampling is based on the correction value and its uncertainty, and obtains the efficiencies distributions with respect to the new correction factors. In the MC simulation model, the uncertainties come from the uncertainties of the amplitude parameters given by the PWA fit to the data. To estimate the uncertainties of the parameters, toy MC samples are generated according to the fitted parameters and the error matrix to obtain the efficiency distribution for each energy point.
The difference in the efficiency determined with pseudoexperiments generated according to the amplitude model parameters with and without the three-body PHSP process is regarded as the uncertainty of the partial wave model.

\par The differences in line shape caused by statistical uncertainties lead to different ISR correction factors. Therefore, for each energy point, the cross section and error serve as the parameters of the Gaussian function of sampling, and after a new round of iteration, the cross section distribution is obtained for each point. Apart from the systematic uncertainty of the ISR correction, the systematic uncertainties of the other contributions are approximately the same for every energy point, and thus the average of each is taken, as listed in Table~\ref{totalsys}. The ISR correction uncertainty of  $R$-scan data samples are estimated by the value of closest $XYZ$ data point, and the uncertainty of each point is listed in the appendix.  Assuming all the sources to be independent, the total systematic uncertainties are obtained by adding them in quadrature.

\begin{table}[!htbp]
\caption{Systematic uncertainties (\%) of the cross section in $\mu^{+}\mu^{-}$ mode and $e^{+}e^{-}$ mode.}
\centering
\renewcommand\arraystretch{1.2}
\renewcommand\tabcolsep{5.0pt}
\begin{tabular}{lll}
\hline
\hline
    \multirow{2}*{Source}               &   \multicolumn{2}{c}{Uncertainty (\%)}                       \\
\cline{2-3}
                                        &       $\mu^{+}\mu^{-}$            &       $e^{+}e^{-}$         \\
\hline
    Luminosities                        &        0.7                       &           0.7             \\
    Tracking efficiency                 &        0.8                       &           0.6             \\
    MC simulation model                            &        1.9                       &           1.9             \\
    Branching fractions                 &        0.6                       &           0.6             \\
    Kinematic fit                       &        0.8                       &           1.3             \\
    Fit to $M(\ell^{+}\ell^{-})$        &        2.4                       &           1.2             \\
    BDT method                          &                                  &          0.9             \\
\hline
    Total                               &        2.7                       &           3.1            \\
\hline
\hline
\end{tabular}
\label{totalsys}
\end{table}

\section{Fit to the cross section}
\label{lineshapes}
\par To study the possible resonant structures in the $\PPJ$ process, a maximum likelihood fit is preformed to the measured cross section. The likelihood is constructed assuming the number of events satisfies a Gaussian distribution in $XYZ$ data and Poisson distribution in $R$-scan data. The cross section is parameterized with a coherent sum of Breit-Wigner (BW) functions. Due to the lack of data near the $\psi(3770)$ resonance, it is not feasible to determine the relative phase between the $\psi(3770)$ amplitude and other amplitudes. The cross section line shape is described by

\begin{equation}\label{ThreeBw}
\footnotesize{
   \sigma_{\mathrm{fit}}(\sqrt{s}) = |R_{\psi(3770)}(\sqrt{s})|^{2} + \left|\sum_{j=0}^{n}R_{j}(\sqrt{s})e^{i\phi_{j}}\right|^{2},
   }
\end{equation}
where $R_{\psi(3770)}$ is used to describe the $\psi(3770)$ resonance and its mass and width are fixed to the world average values~\cite{PDG}. The $i$ is the imaginary unit. $R_{j}$ represents the amplitude to describe a given resonant structure and $\phi_{j}$ is the corresponding phase. The phase $\phi_{0}$ is set to zero and the other phases are given relative to the $R_{0}$. For the structure near 4.0 GeV, two different parameterization methods are applied, Model I: a BW function, and Model II: an exponential function (Exp) of the form $R_{0}(\sqrt{s}) = PS(\sqrt{s})e^{-p_{0}(\sqrt{s}-M_{\rm{threshold}})}p_{1}$~\cite{STAR:2018fty}, with $M_{\rm {threshold}} = m_{\pi^{+}}+m_{\pi^{-}}+m_{J/\psi}$, $PS(\sqrt{s})$ is the PHSP factor of the three-body decay $R_{j} \rightarrow \pi^{+}\pi^{-}J/\psi$~\cite{PDG}, and $p_{0}$ and $p_{1}$ are free parameters determined by the fit. The number of resonances is denoted by $n$, comprising the known $Y(4220)$ and $Y(4320)$ as well as further possible structures. The amplitude $R_{j}$ is defined as

\begin{equation}\label{fiF}
\footnotesize{
  R_{j}(\sqrt{s}) = \frac{M_{j}}{\sqrt{s}} \frac{\sqrt{12\pi \Gamma^{\rm ee}_{j}\Gamma^{\rm tot}_{j}\mathcal{B}(R_{j})}}{s-M^{2}_{j}+iM_{j}\Gamma^{\rm{tot}}_{j}} \times \sqrt{\frac{PS(\sqrt{s})}{PS(M_{j})}},
}
\end{equation}
where $M_{j}$, $\Gamma^{\rm tot}_{j}$ and $\Gamma^{\rm ee}_{j}$ are the mass, full width and electronic width of resonance $R_{j}$, respectively, and $\mathcal{B}(R_{j})$ is the branching fraction for $R_{j}\rightarrow \pi^{+}\pi^{-}J/\psi$.

\par
\begin{figure*}
\centering
  \begin{minipage}{1.0\linewidth}
  \includegraphics[width=0.49\textwidth, height=0.45\textwidth]{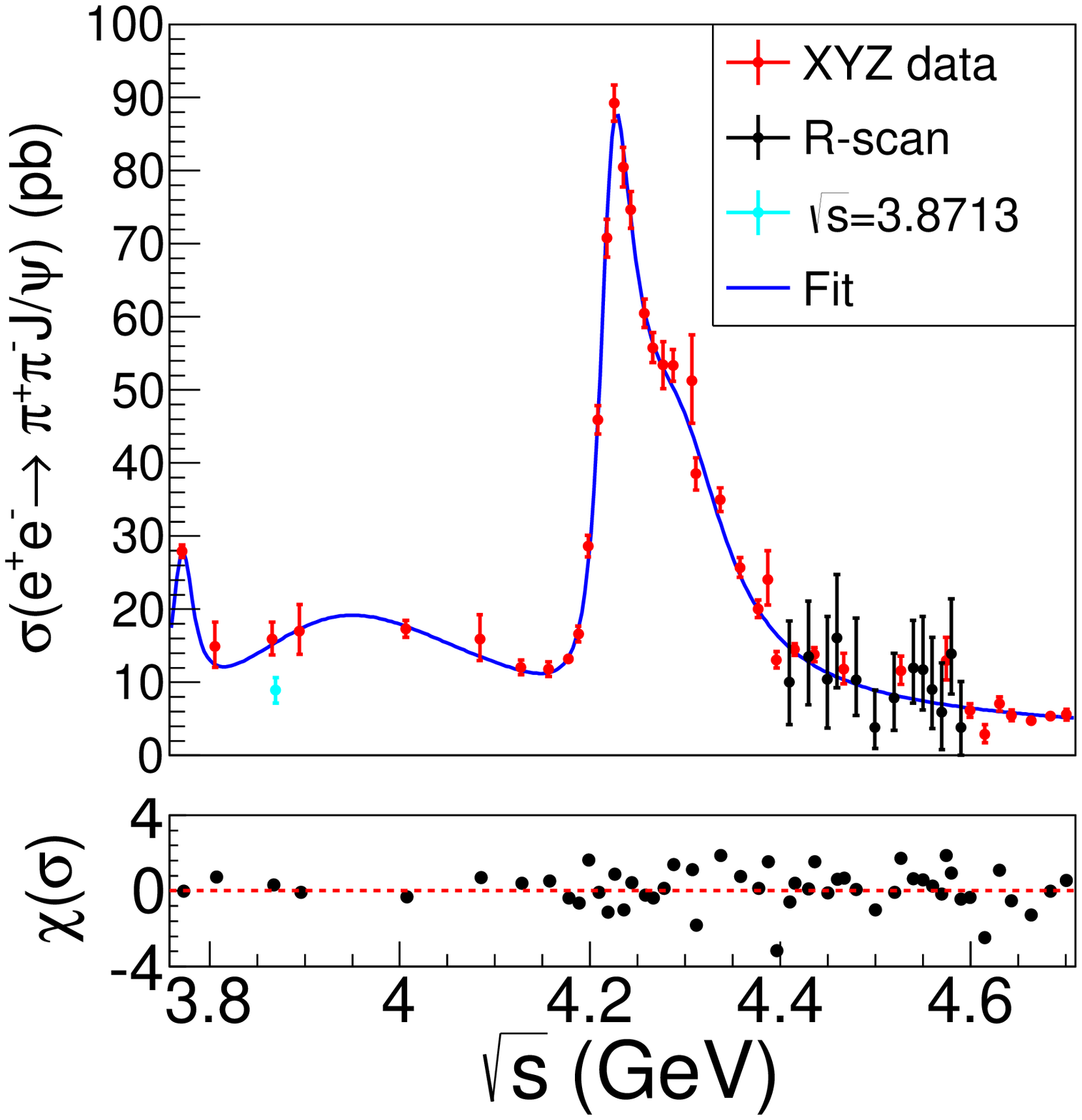}
  \includegraphics[width=0.49\textwidth, height=0.45\textwidth]{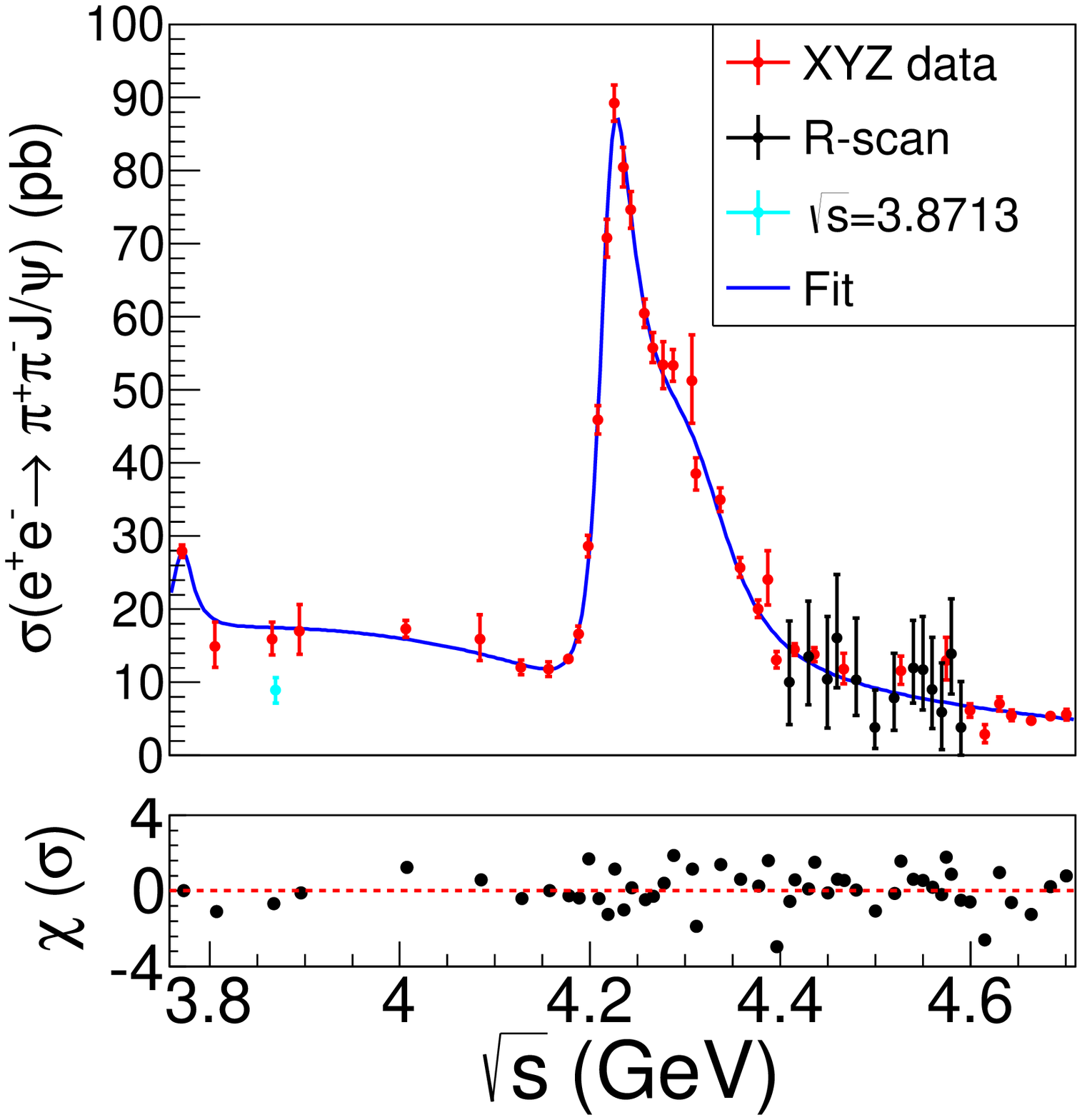}
  \end{minipage}
  \put(-375, 62){(a)}
  \put(-160, 62){(b)}
\caption{\footnotesize{Fit to the energy-dependent cross section of $e^{+}e^{-} \rightarrow  \pi^{+}\pi^{-}J/\psi$ using two different fit models: Model I (a) and Model II (b). The upper panels show the data points with error bars overlaid with the fit result represented by the solid (blue) line. The lower panels show the corresponding fit quality for each data point in terms of $\chi$ in units of $\sigma$. The point of $\sqrt{s} = 3.8713$ GeV is not included in the fit. For more details of the fit models, see the text.}}
\label{FitCsLike1}
\end{figure*}

 In case of considering the states $Y(4220)$ and $Y(4320)$ ($n=2$), multiple sets of solutions are obtained given by the two models (Model I: BW, Model II: Exp) at 4.0 GeV. The fit results are shown in Fig.~\ref{FitCsLike1}, and the fit parameters are summarized in Table~\ref{Result3BW}. Sizable differences between the fit results of Model I and Model II appear mainly in the energy region between 3.7730 and 4.1574 GeV. The difference of $\chi^{2}/\rm{ndf}$ is 3.72, where $\rm{ndf}$ is the number of degrees of freedom. Therefore, Model I is chosen to be the default model for the final cross section fit result. The cross section fit shows larger fluctuations at $\sqrt{s} =$ 3.8713 GeV. These might be due to the influence of the $X(3872)$~\cite{PDG} resonance which was not included in the model since the $X(3872)$ is very narrow and there are not sufficient data points around its nominal mass.

\begin{table*}
\caption{\footnotesize{The values of $\Gamma^{\rm e^{+}e^{-}}\mathcal{B}(R \rightarrow \pi^{+}\pi^{-} J/\psi)$ from the fit to the $\PPJ$ cross section. The parameters $M(R_{n})$, $\Gamma^{\rm tot}_{n}(R_{n})$ and $\Gamma^{\rm ee}_{n}\mathcal{B}(R_{n})$, ($n =0, 1, 2$) represent the mass (in MeV/$c^{2}$), total width (in MeV) and the product of the $e^{+}e^{-}$ partial width (in eV) with the branching ratio of the resonance decay into $\pi^{+}\pi^{-}J/\psi$ for the resonances, respectively. Here, $p_{0}$ ($c^{2}$/MeV) and $p_{1}$ are the free parameters of the exponential function. The parameters $\phi_{1}$ and $\phi_{2}$ (in degrees) are phases of the resonance $R_{1}$ and $R_{2}$, the phase of resonance $R_{0}$ is set to 0. The numbers in the brackets correspond to the fit by replacing the resonance $R_{0}$ (BW: Model I) with an exponential function (Exp: Model II) to describe the structure near 4.0 GeV. The uncertainties are statistical only. ``..." represents a null value.}}
\centering
\resizebox{\textwidth}{!}{
\renewcommand\arraystretch{1.4}
\renewcommand\tabcolsep{5.0pt}
\begin{tabular}{lrrrr}
\hline
\hline
  Parameter                          &   Solution I       &   Solution II    & Solution III     & Solution IV      \\
\hline
  $\Gamma^{\rm ee}_{3770}\mathcal{B}(R_{3770})$ &                     \multicolumn{4}{c}{$0.6\pm0.1~(0.3\pm0.1)$}   \\
  $M(R_{0})$ ($p_{0}$)               & \multicolumn{4}{c}{$3905.5\pm30.1$~($4.4\pm0.3$)}                                 \\
  $\Gamma^{\rm tot}_{0}(R_{0})$ ($p_{1}$)    & \multicolumn{4}{c}{$346.0\pm48.5$~($(2.7\pm0.6)\times10^{-3}$)}      \\
\hline
  $\Gamma^{\rm ee}_{0}\mathcal{B}(R_{0})$    & $5.5\pm0.5$~(...) & $6.9\pm0.7$~(...) & $8.3\pm0.6$~(...) & $10.5\pm0.9$~(...) \\
\hline
  $M(R_{1})$                         & \multicolumn{4}{c}{$4221.4\pm1.5$~($4220.1\pm1.2$)}    \\
  $\Gamma^{\rm tot}_{1}(R_{1})$              & \multicolumn{4}{c}{$41.8\pm2.9$~($43.6\pm2.6$)}       \\
\hline
  $\Gamma^{\rm ee}_{1}\mathcal{B}(R_{1})$    & $1.7\pm0.2~(1.7\pm0.2)$ & $8.2\pm0.9~(8.6\pm0.5)$ & $3.0\pm0.5~(2.5\pm0.3)$ & $14.6\pm1.2~(12.7\pm0.8)$  \\
\hline
  $M(R_{2})$                         & \multicolumn{4}{c}{$4297.5\pm12.1$~($4316.2\pm12.4$)}  \\
  $\Gamma^{\rm tot}_{2}(R_{2})$              & \multicolumn{4}{c}{$126.6\pm16.7$~($124.3\pm18.0$)}   \\
\hline
  $\Gamma^{\rm ee}_{2}\mathcal{B}(R_{2})$    & $1.2\pm0.3~(0.7\pm0.2)$ & $2.3\pm0.8~(1.1\pm0.3)$ & $15.6\pm2.1~(15.0\pm1.2)$    & $30.2\pm3.3~(23.6\pm2.9)$    \\
  $\phi_{1}$                         & $-3.7\pm5.4~(24.3\pm3.0)$  & $-124.6\pm11.7~(-78.8\pm5.1)$& $87.7\pm21.9~(88.0\pm12.1)$ & $-33.5\pm11.2~(-15.1\pm7.7)$  \\
  $\phi_{2}$                         & $79.6\pm18.5~(130.7\pm15.8)$ & $35.8\pm27.2~(96.6\pm19.7)$  & $-104.7\pm26.9~(-92.5\pm6.0)$ & $-148.7\pm4.5~(-127.6\pm2.3)$   \\
\hline
  $\chi^{2}/\rm{ndf}$   & \multicolumn{4}{c}{ 54.0/40 (57.3/41)} \\
\hline
\hline
\end{tabular}}
\label{Result3BW}
\end{table*}

\par Considering the distribution of the pull ($\chi$) values, the above two models do not describe the interval from 4.4 to 4.6\ GeV very well. Therefore, a third BW function ($n=3$) is added to study whether this deviation is caused by possible additional structures. When the (fit) parameters of the third BW function are floated, two possible solutions are obtained, one with a mass close to the $\psi(4415)$ and the other one close to 4.5\ GeV. Compared with the mode of $n=2$, the significance of these two solutions are $4.0\sigma$ ($3.6\sigma$) and $2.1\sigma$ ($2.7\sigma$), respectively. The alternative fits using the parameters of the $\psi(4415)$~\cite{PDG} and the newly discovered $Y(4500)$ structure~\cite{BESIII:2022joj} have also been attempted to describe the structure at 4.5\ GeV, and led to results with significance of $2.6\sigma$ ($3.1\sigma$) and $3.3~\sigma$ ($3.3~\sigma$), respectively. The numbers in the brackets correspond to an alternative fit, in which the BW function(Model I) is replaced by an exponential function(Model II) to describe the structure near 4.0\ GeV, as also shown in Table~\ref{Result3BW} and Table~\ref{Result4BW}.

\par
\begin{figure*}
\centering
  \begin{minipage}{1.0\linewidth}
    \includegraphics[width=0.49\textwidth, height=0.45\textwidth]{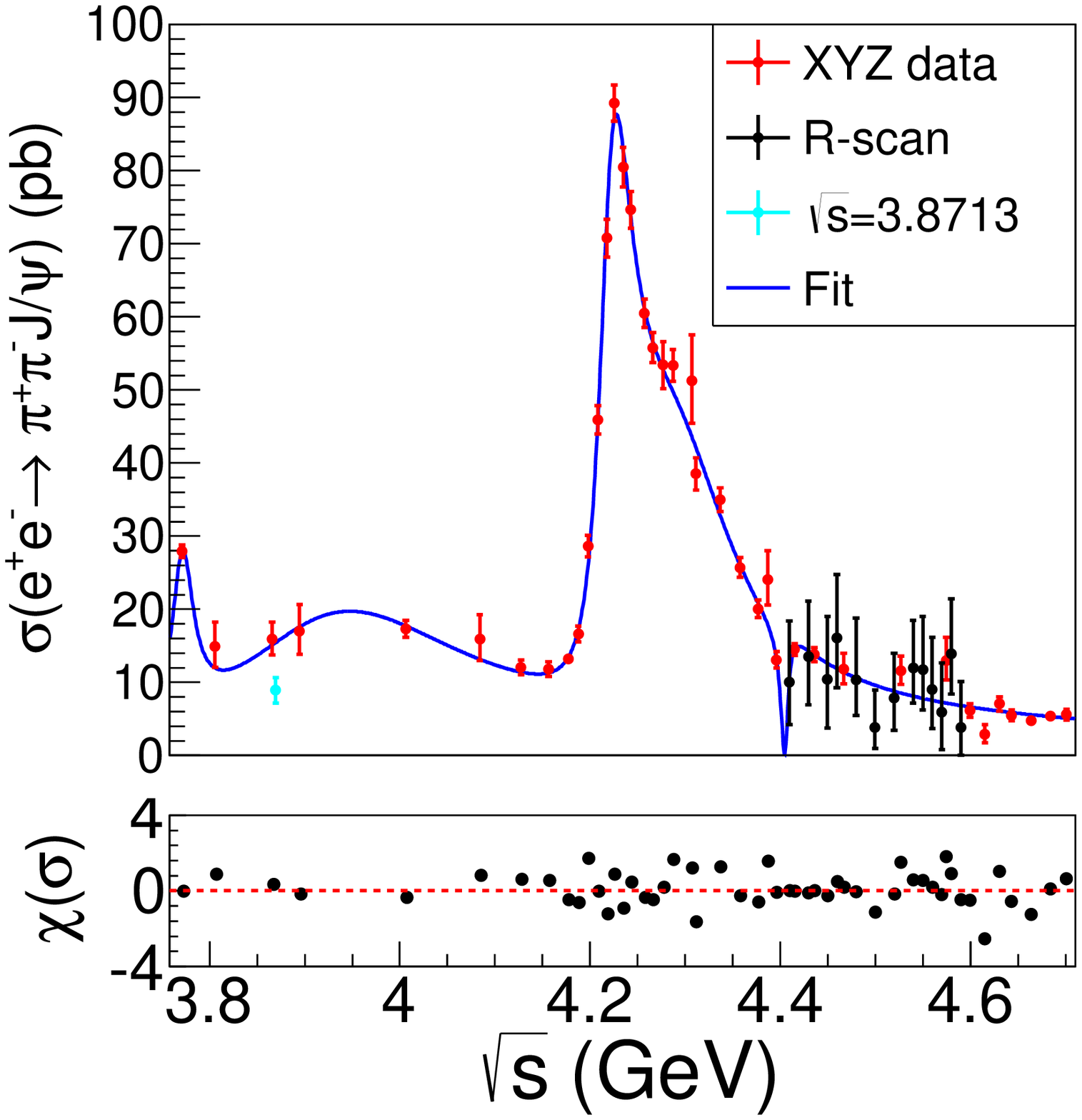}
    \includegraphics[width=0.49\textwidth, height=0.45\textwidth]{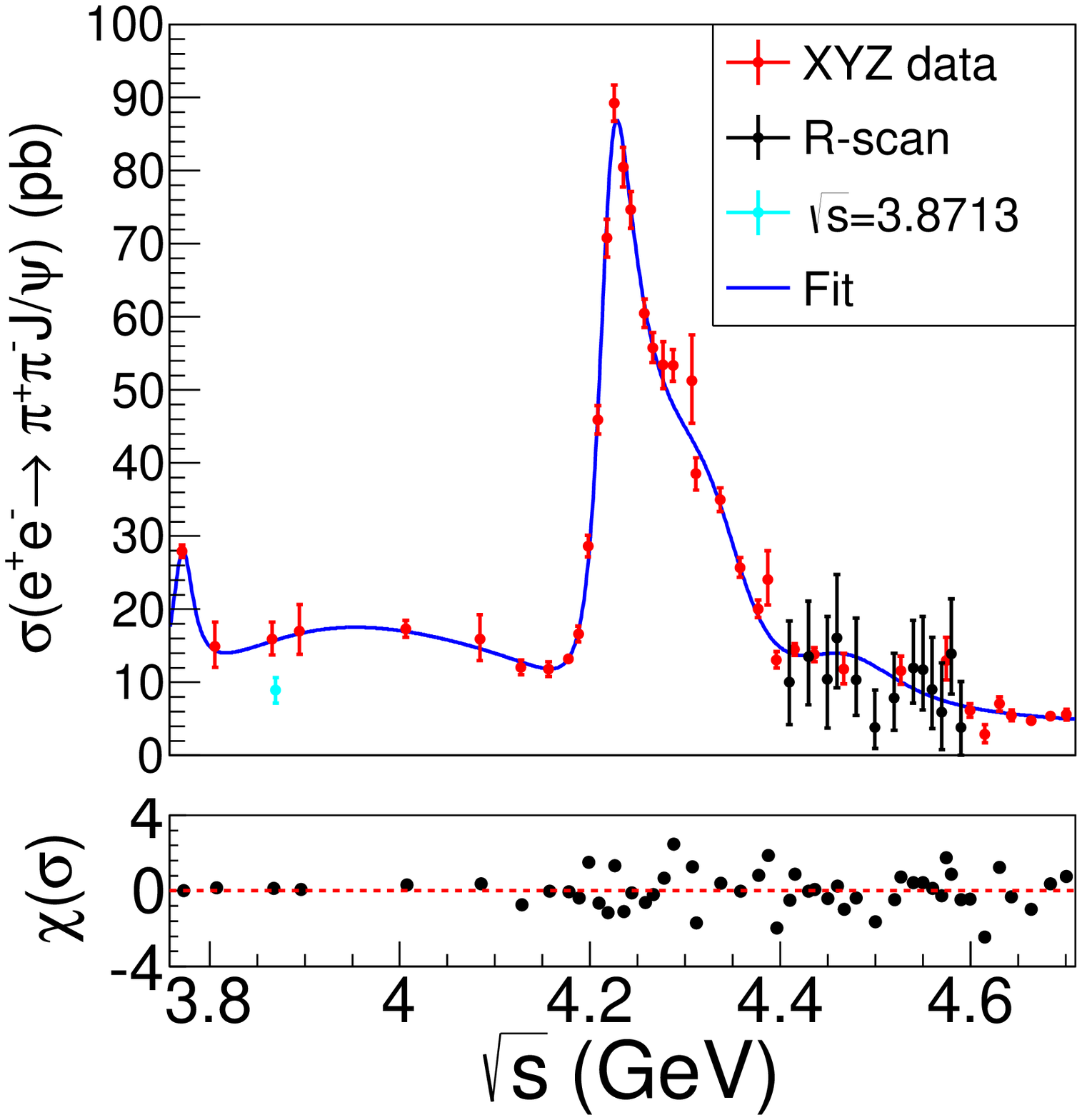}
    \put(-310, -2){(a)}
    \put(-100, -2){(b)}
  \end{minipage}

  \begin{minipage}{1.0\linewidth}
    \includegraphics[width=0.49\textwidth, height=0.45\textwidth]{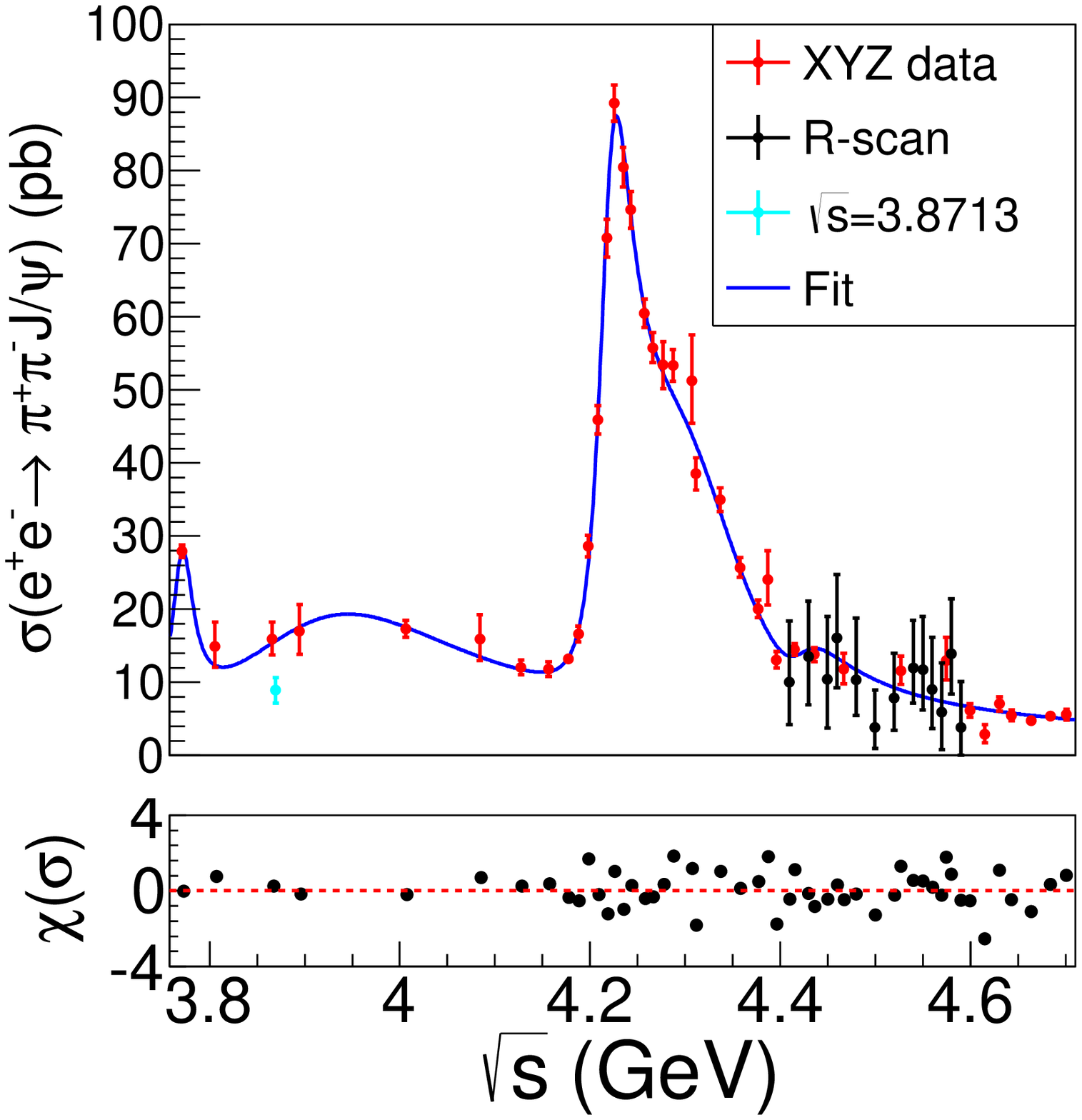}
    \includegraphics[width=0.49\textwidth, height=0.45\textwidth]{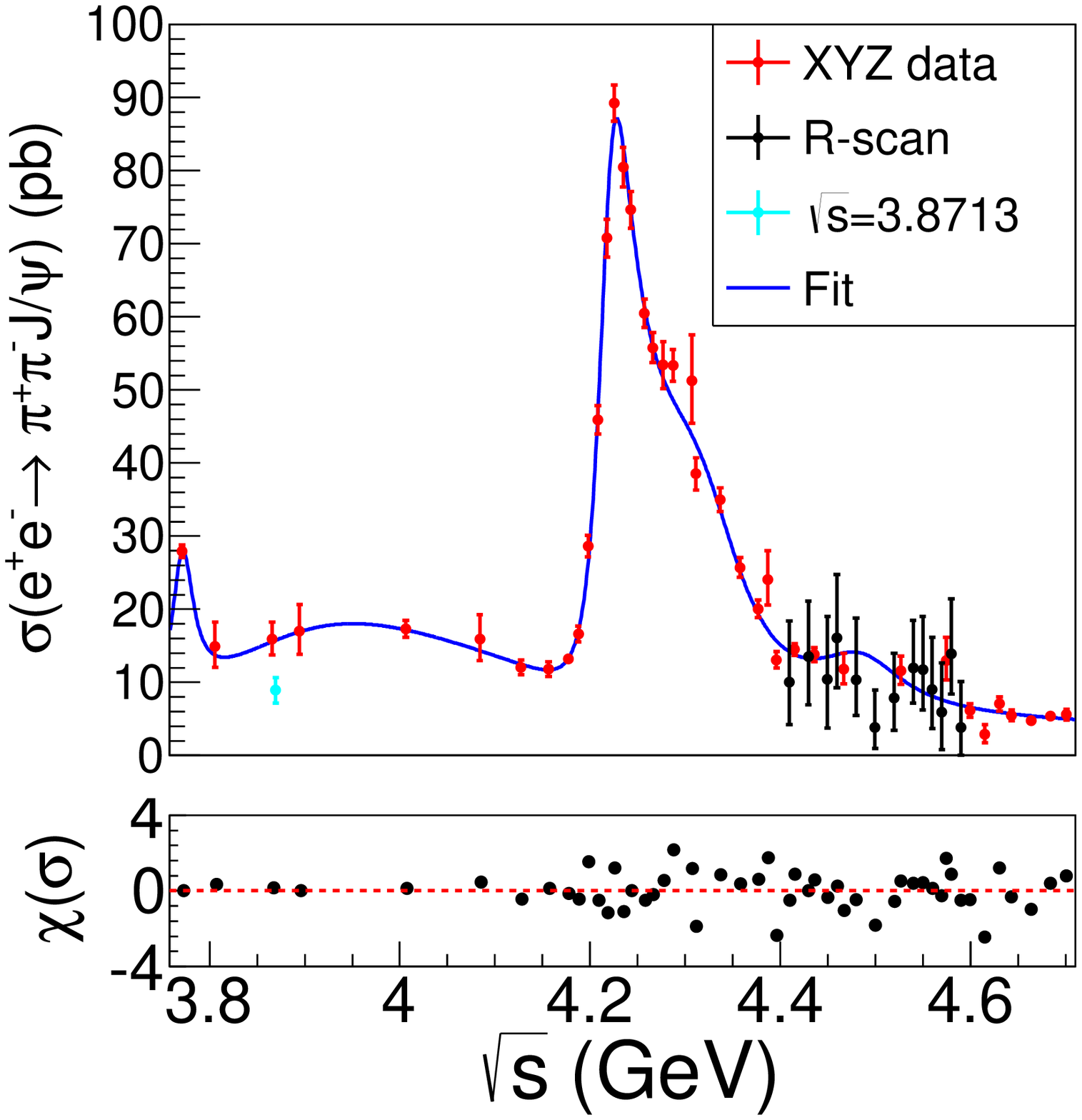}
  \put(-310, -2){(c)}
  \put(-100, -2){(d)}
  \end{minipage}
\caption{\footnotesize{Fit to the energy-dependent cross section of the process $e^{+}e^{-} \rightarrow  \pi^{+}\pi^{-}J/\psi$, where an additional BW function is added based on Model I. The upper panels show the data points with error bars overlaid with the fit result represented by the solid (blue) line. The lower panels show the corresponding fit quality for each data point in terms of $\chi$ in units of $\sigma$, and the fit results are listed in Table~\ref{Result4BW}. (a) and (b): The new BW function with free parameters. (c) and (d): The new BW function with parameters fixed to the $\psi(4415)$ and $Y(4500)$ resonance parameters. The point of $\sqrt{s} = 3.8713$ GeV is not included in the fit.}}
\label{FitCsLike_4BW1}
\end{figure*}

 Due to the limited data samples around 4.4\ GeV, the mass and width of the new additional BW function are fixed to $\psi(4415)$ or $Y(4500)$, the fit result in Table~\ref{Result4BW}. With larger/new data samples in that region, the structure of the $\psi(4415)$ or $Y(4500)$ can be studied further. The fit results are shown in Fig.~\ref{FitCsLike_4BW1}, and the parameters are summarized in Table~\ref{Result4BW}. In conclusion, the parameters of the $Y(4220)$ state are stable for the different models, while the ones of $Y(4360)$ heavily depend on the presence of an additional structure close to 4.5\ GeV.

\begin{table*}
\caption{\footnotesize{The fit results of the test for the additional structure near 4.5~GeV. The values of $\Gamma^{\rm e^{+}e^{-}}\mathcal{B}(R \rightarrow \pi^{+}\pi^{-} J/\psi)$ from the fit to the $\PPJ$ cross section. The parameters $M(R_{n})$, and $\Gamma^{\rm tot}_{n}(R_{n})$ ($n =0, 1, 2, 3$) represent the mass (in MeV/$c^{2}$), total width (in MeV) of the resonance decay into $\pi^{+}\pi^{-}J/\psi$ for the resonances, respectively. Finally, the significance of the additional BW is given. The uncertainties are statistical only. The numbers in the brackets correspond to the fit by replacing the BW:Model I with an Exp: Model II to describe the structure near 4.0 GeV.}}
\centering
\resizebox{\textwidth}{!}{
\renewcommand\tabcolsep{10.0pt}
\renewcommand\arraystretch{1.2}
\footnotesize
\begin{tabular}{lllll}
\hline
\hline
  Parameter          &   Result I       &   Result II      & Result III    & Result IV   \\
\hline
  $M(R_{1})$                         &  $4221.0\pm1.6$~($4220.3\pm1.6$)    &  $4219.8\pm1.3$~($4219.1\pm1.2$) & $4223.9\pm1.4$~($4219.6\pm1.3$) & $4220.2\pm1.3$~($4219.4\pm1.1$) \\
  $\Gamma^{\rm tot}_{1}(R_{1})$          &  $41.0\pm3.0$~($42.3\pm3.0$)        &  $45.4\pm2.8$~($46.3\pm2.5$) & $42.2\pm3.2$~($44.3\pm2.7$) & $44.5\pm2.9$~($45.3\pm2.5$)     \\
  $M(R_{2})$                         &  $4293.7\pm13.1$~($4304.8\pm18.8$)  &  $4345.8\pm28.4$~($4357.9\pm20.2$) &  $4308.5\pm17.6$~($4333.2\pm23.2$)  & $4328.58\pm18.9$~($4347.1\pm14.5$)  \\
  $\Gamma^{\rm tot}_{2}(R_{2})$          &  $152.4\pm23.9$~($144.3\pm31.5$)    &  $130.1\pm20.7$~($107.9\pm25.6$) & $161.4\pm24.6$~($153.2\pm26.2$)  & $133.8\pm20.2$~($127.5\pm22.2$)\\
  $M(R_{3})$                         &  $4405.6\pm4.5$~($4405.0\pm6.7$)    &  $4471.1\pm36.2$~($4550.9\pm16.9$) & 4421~(fixed) & 4485~(fixed)\\
  $\Gamma^{\rm tot}_{3}(R_{3})$          &  $9.1\pm2.5$~($8.7\pm4.9$)          &  $159.7\pm97.0$~($211.8\pm132.8$) &  62~(fixed)  & 111~(fixed) \\
\hline
  $\chi^{2}/\rm{ndf}$   &    40.1/36 (44.8/37)   & 47.6/36 (48.7/37) & 45.4/38 (48.7/39)  & 48.1/38 (51.3/39)\\
  Significance                       &    $4.0\sigma$~($3.6\sigma$)    &   $2.1\sigma$~($2.7\sigma$)    &  $3.3\sigma$ ~($3.3\sigma$)  & $2.6\sigma$ ~($3.1\sigma$)   \\
\hline
\hline
\end{tabular}}
\label{Result4BW}
\end{table*}

\section{Systematic uncertainties of the resonance parameters}
\par Relevant systematic uncertainties of the fit parameters are caused by the fit model, the $\sqrt{s}$ energy measurement and its spread, and the PHSP factor. The systematic uncertainty related to the fit model is evaluated as the difference of the mass and width results of Model I and Model II. The $\sqrt{s}$ of all data sets have been measured with di-muon events with an uncertainty of 0.6 MeV that propagates directly to the uncertainty of the mass of the resonances. The uncertainties included by the $\sqrt{s}$ spread are obtained by convolving the resonant PDF with a Gaussian function whose width is taken to be 1.6 MeV, equal to the spread obtained from the Beam Energy Measurement System~\cite{Abakumova:2011rp}. The uncertainty of the PHSP factor, due to the existence of intermediate states, is estimated by considering the PHSP of cascade two-body decays of $\EE \rightarrow R J/\psi$ (with $R = \sigma,~f_{0}(980),~f_{0}(1370)$) and $\EE \rightarrow \pi^{\pm}Z_{c}(3900)^{\mp}$ , and the maximum value of the difference with respect to the result obtained when using the three-body PHSP factor is taken as the systematic uncertainty. The deviation of the resonant parameters introduced by the uncertainties of the $\psi(3770)$ resonance parameters are less than 0.1 MeV, and thus can be neglected. Assuming all of the systematic uncertainties are independent, adding them in quadrature delivers the total error as listed in Table~\ref{Csfitsys}.

\begin{table}[!htbp]
\caption{Summary of the uncertainties of the resonance parameters.}
\centering
\scalebox{0.7}{
\begin{tabular}{lcccc}
\hline
\hline
    \multirow{3}*{Source}      &   \multicolumn{4}{c}{Uncertainty}                       \\
\cline{2-5}
                               &   \multicolumn{2}{c}{$Y(4220)$}   &    \multicolumn{2}{c}{$Y(4320)$}    \\
                               &    M (MeV/$c^{2}$)  &   $\Gamma$ (MeV) &  M (MeV/$c^{2}$)  &  $\Gamma$ (MeV)     \\
\hline
    $\sqrt{s}$                 &    0.6   &                      &    0.6      &                     \\
    Beam spread                &    0.3   &   0.4                &    5.0      &           2.1       \\
    Fit model                  &    1.4   &   1.0                &    15.8     &           6.8       \\
    PHSP factor                &    1.3   &   2.5                &    19.9     &           7.8      \\
\hline
    Total                      &    2.0   &   2.7                &    25.9     &           10.3      \\
\hline
\hline
\end{tabular}}
\label{Csfitsys}
\end{table}

\section{Summary}
\par In summary, a precision measurement of the energy-dependent cross section for the process $e^{+}e^{-} \rightarrow \pi^{+}\pi^{-}J/\psi$ from $\sqrt{s} = 3.7730$ GeV to $\sqrt{s} = 4.7008$ GeV at BESIII is performed. This measurement improves upon the precision of the previous results from the BESIII collaboration in the same channel~\cite{BESIII:2016bnd} by about 60\% at the points of same statistics ($\sqrt{s} = 4.2263$ and 4.2580 GeV). This is achieved by the improvements in the optimized MC simulation model and the enhanced tracking efficiencies.

\par The energy-dependent cross section is fitted with different fit models for the cross section line shape, allowing for the search of resonances and the evaluation of their parameters. It was found that the structure close to 4.0 GeV is better described when using the BW function as compared to an Exp function. The $Y(4220)$ and $Y(4320)$ resonances were observed with significances larger than $10\sigma$ and their resonance parameters were estimated to be consistent with those reported in Ref.~\cite{BESIII:2016bnd}. However, the presence of an additional structure around 4.5~GeV, possibly identifiable with the $\psi(4415)$, influences the evaluation of the $Y(4320)$ parameters, which are ($M,~\Gamma$) = ($4298\pm12\pm26$\ MeV/$c^{2}$, $127\pm17\pm10$\ MeV), therefore, reported with a large uncertainty.

\acknowledgements

The BESIII collaboration thanks the staff of BEPCII and the IHEP computing center for their strong support. This work is supported in part by National Key R\&D Program of China under Contracts Nos. 2020YFA0406300, 2020YFA0406400; National Natural Science Foundation of China (NSFC) under Contracts Nos. 11625523, 11635010, 11735014, 11822506, 11835012, 11935015, 11975141, 11935016, 11935018, 11961141012, 12022510, 12025502, 12035009, 12035013, 12061131003, 11875262; the Chinese Academy of Sciences (CAS) Large-Scale Scientific Facility Program; Joint Large-Scale Scientific Facility Funds of the NSFC and CAS under Contracts Nos. U1732263, U1832207; CAS Key Research Program of Frontier Sciences under Contract No. QYZDJ-SSW-SLH040; 100 Talents Program of CAS; INPAC and Shanghai Key Laboratory for Particle Physics and Cosmology; ERC under Contract No. 758462; European Union Horizon 2020 research and innovation programme under Contract No. Marie Sklodowska-Curie grant agreement No 894790; The Fundamental Research Funds of Shandong University; German Research Foundation DFG under Contracts Nos. 443159800, Collaborative Research Center CRC 1044, FOR 2359, GRK 2149; Istituto Nazionale di Fisica Nucleare, Italy; Ministry of Development of Turkey under Contract No. DPT2006K-120470; National Science and Technology fund; Olle Engkvist Foundation under Contract No. 200-0605; STFC (United Kingdom); The Knut and Alice Wallenberg Foundation (Sweden) under Contract No. 2016.0157; The Royal Society, UK under Contracts Nos. DH140054, DH160214; The Swedish Research Council; U. S. Department of Energy under Contracts Nos. DE-FG02-05ER41374, DE-SC-0012069.



\section*{Appendix}
\par Table~\ref{totCrosssection} lists the parameters used to calculate the cross section of the $\PPJ$ process with the Eq.~\ref{Cscalculate}.

\newpage

\thispagestyle{empty}


\begin{table*}
\caption{The cross section of $e^{+}e^{-}\rightarrow \pi^{+}\pi^{-}J/\psi$. Here, $\mathcal{L}$ (pb$^{-1}$) is the integrated luminosity, $1+\delta$ is the radiation correction factor, ``Rsys" is the systematic uncertainties (\%) of the radiation correction factor, N$^{\rm obs}$ is the number of observed signal events from $\ell^{+}\ell^{-}$ invariant mass distributions, $\epsilon$ is the event selection efficiency determined from signal MC, $\sigma$ (pb) is the cross section. ``average" is the weighted average cross section of $\EE$ and $\MM$ modes, and the weight is the inverse of the statistical uncertainty. Samples marked with ``$r$" are $R$-scan data, and the results are given by combined $e^{+}e^{-}$ mode and $\mu^{+}\mu^{-}$ mode. The first term is statistical uncertainty, and the second term is systematic uncertainty. The uncertainty of $\mathcal{L}$ consists of statistical and systematic uncertainties. The uncertainty for N$^{\rm obs}$ is statistical only.}
\centering
\renewcommand\arraystretch{1.1}
\renewcommand\tabcolsep{7.0pt}
\scalebox{0.5}{
\begin{tabular}{lrrrrrrrrrr}
\hline
\hline
  \multirow{2}*{ $\sqrt{s}$ (GeV)} & \multirow{2}*{$\mathcal{L}$ (pb$^{-1}$)} & \multirow{2}*{$1+\delta$} & \multirow{2}*{Rsys(\%)} & \multicolumn{2}{c}{N$^{\rm obs}$} & \multicolumn{2}{c}{$\epsilon$ (\%)} & \multicolumn{3}{c}{$\sigma$ (pb)}  \\
                 &          &        &      & \makecell[c]{$e^{+}e^{-}$} &   \makecell[c]{$\mu^{+}\mu^{-}$}   & \makecell[c]{ $e^{+}e^{-}$} &  \makecell[c]{$\mu^{+}\mu^{-}$ }  & \makecell[c]{ $e^{+}e^{-}$ } & \makecell[c]{ $\mu^{+}\mu^{-}$ }  & \makecell[c]{ average} \\
\hline
  3.7730 &$2874.72\pm19.40$& 0.73 & 2.04 & $1169\pm36$ & $1756\pm46$ &  32.41 & 51.34 & $28.74\pm0.89\pm1.06$ & $27.30\pm0.72\pm0.92$  &$27.94\pm0.56\pm0.69$ \\
  3.8077 & $50.54\pm0.49$  & 0.86 & 10.46 & $12\pm4$    & $17\pm5$ &  29.60 & 46.21 & $15.67\pm5.22\pm1.71$ & $14.26\pm4.19\pm1.54$  &$14.89\pm3.29\pm1.14$ \\
  3.8674 & $108.90\pm0.72$ & 0.86 & 2.30  & $24\pm5$    & $40\pm8$ &  27.09 & 44.39 & $15.77\pm3.28\pm0.61$ & $16.05\pm3.21\pm0.57$  &$15.91\pm2.30\pm0.41$ \\
  3.8713 & $110.30\pm0.73$ & 1.19 & 14.29 & $17\pm5$    & $30\pm6$ &  24.96 & 41.99 & $8.67\pm2.55\pm1.27$  & $9.10\pm1.82\pm1.32$   &$8.92\pm1.50\pm0.93$ \\
  3.8962 & $52.61\pm0.51$  & 0.86 & 5.00  & $14\pm4$    & $19\pm5$ &  27.39 & 45.41 & $18.93\pm5.41\pm1.11$ & $15.52\pm4.08\pm0.88$  &$16.98\pm3.29\pm0.69$ \\
  4.0076 & $482.00\pm3.18$ & 0.90 & 1.15  & $120\pm12$  & $229\pm18$  &  29.63 & 46.71 & $15.62\pm1.56\pm0.51$ & $18.94\pm1.49\pm0.55$  &$17.32\pm1.08\pm0.38$ \\
  4.0855 & $52.86\pm0.35$  & 0.92 & 3.28  & $15\pm4$    & $20\pm5$    &  29.65 & 47.00 & $17.43\pm4.65\pm0.78$ & $14.68\pm3.67\pm0.62$  &$15.89\pm2.90\pm0.49$ \\
  4.1271 & $397.99\pm2.63$ & 0.97 & 3.19  & $86\pm10$   & $127\pm14$  &  29.66 & 47.37 & $12.52\pm1.46\pm0.55$ & $11.60\pm1.28\pm0.48$  &$12.03\pm0.96\pm0.36$ \\
  4.1567 & $409.88\pm2.70$ & 0.96 & 2.15  & $83\pm10$   & $118\pm13$  &  28.08 & 44.98 & $12.61\pm1.52\pm0.47$ & $11.21\pm1.24\pm0.38$  &$11.84\pm0.96\pm0.30$ \\
  4.1780 &$3194.50\pm31.90$& 0.94 & 2.61  & $612\pm27$  & $1032\pm38$ &  26.48 & 42.84 & $12.87\pm0.57\pm0.52$ & $13.44\pm0.49\pm0.50$  &$13.17\pm0.37\pm0.36$ \\
  4.1888 & $570.05\pm2.16$ & 0.90 & 2.83  & $150\pm13$  & $231\pm19$  &  28.85 & 46.23 & $16.94\pm1.47\pm0.71$ & $16.31\pm1.34\pm0.64$  &$16.61\pm0.99\pm0.47$ \\
  4.1989 & $524.60\pm2.05$ & 0.81 & 2.49  & $242\pm17$  & $354\pm21$  &  31.93 & 50.14 & $29.72\pm2.09\pm1.17$ & $27.77\pm1.65\pm1.02$  &$28.63\pm1.30\pm0.77$ \\
  4.2091 & $572.05\pm1.81$ & 0.79 & 2.26  & $389\pm21$  & $634\pm27$  &  32.24 & 49.92 & $44.56\pm2.41\pm1.70$ & $47.07\pm2.00\pm1.65$  &$45.93\pm1.55\pm1.19$ \\
  4.2187 & $569.20\pm1.80$ & 0.77 & 1.94  & $627\pm26$  & $915\pm32$  &  32.95 & 50.88 & $72.90\pm3.02\pm2.65$ & $69.06\pm2.42\pm2.29$  &$70.77\pm1.90\pm1.73$ \\
  4.2263 & $1100.91\pm7.00$& 0.76 & 1.15  & $1517\pm40$ & $2315\pm51$ &  34.37 & 52.00 & $88.68\pm2.34\pm2.88$ & $89.81\pm1.98\pm2.59$  &$89.29\pm1.52\pm1.93$ \\
  4.2357 & $530.60\pm2.39$ & 0.82 & 1.54  & $696\pm27$  & $1103\pm35$ &  34.05 & 51.69 & $78.50\pm3.05\pm2.70$ & $82.20\pm2.61\pm2.55$  &$80.50\pm1.99\pm1.85$ \\
  4.2438 & $593.98\pm2.69$ & 0.84 & 1.42  & $744\pm28$  & $1133\pm36$ &  33.22 & 51.12 & $74.99\pm2.82\pm2.54$ & $74.37\pm2.36\pm2.26$  &$74.65\pm1.82\pm1.69$ \\
  4.2580 & $828.40\pm5.47$ & 0.89 & 1.15  & $888\pm31$  & $1327\pm39$ &  32.77 & 50.17 & $61.25\pm2.14\pm2.01$ & $59.89\pm1.76\pm1.75$  &$60.50\pm1.37\pm1.32$ \\
  4.2667 & $529.70\pm3.13$ & 0.91 & 1.43  & $509\pm23$  & $781\pm30$  &  31.73 & 49.14 & $56.03\pm2.53\pm1.90$ & $55.56\pm2.13\pm1.69$  &$55.78\pm1.64\pm1.26$ \\
  4.2776 & $175.70\pm0.96$ & 0.91 & 2.87  & $152\pm13$  & $253\pm17$  &  31.01 & 48.02 & $51.29\pm4.39\pm2.16$ & $55.20\pm3.71\pm2.17$  &$53.41\pm2.84\pm1.54$ \\
  4.2866 & $498.52\pm3.29$ & 0.90 & 2.32  & $451\pm22$  & $684\pm28$  &  31.34 & 47.82 & $53.53\pm2.61\pm2.06$ & $53.27\pm2.18\pm1.89$  &$53.40\pm1.68\pm1.39$ \\
  4.3079 & $45.08\pm0.30$  & 0.87 & 4.04  & $35\pm6$    & $59\pm9$    &  30.94 & 47.07 & $48.41\pm8.30\pm2.46$ & $53.74\pm7.29\pm2.61$  &$51.25\pm5.49\pm1.79$ \\
  4.3115 & $499.19\pm3.30$ & 1.05 & 5.74  & $345\pm19$  & $573\pm26$  &  29.99 & 45.90 & $36.85\pm2.03\pm2.40$ & $40.03\pm1.82\pm2.54$  &$38.53\pm1.36\pm1.75$ \\
  4.3370 & $511.48\pm3.38$ & 0.98 & 1.86  & $313\pm19$  & $453\pm24$  &  28.85 & 44.30 & $36.08\pm2.19\pm1.30$ & $34.06\pm1.80\pm1.11$  &$34.97\pm1.40\pm0.85$ \\
  4.3583 & $543.94\pm3.59$ & 1.06 & 2.98  & $267\pm17$  & $342\pm21$  &  27.01 & 42.17 & $28.55\pm1.82\pm1.22$ & $23.47\pm1.44\pm0.94$  &$25.72\pm1.14\pm0.75$ \\
  4.3768 & $531.41\pm3.51$ & 1.14 & 3.41  & $195\pm15$  & $297\pm20$  &  26.46 & 41.38 & $20.32\pm1.56\pm0.93$ & $19.83\pm1.34\pm0.86$  &$20.06\pm1.02\pm0.63$ \\
  4.3874 & $55.57\pm0.37$  & 0.98 & 4.93  & $22\pm5$    & $32\pm6$    &  27.14 & 41.77 & $24.82\pm5.64\pm1.44$ & $23.51\pm4.41\pm1.32$  &$24.09\pm3.50\pm0.97$ \\
  4.3954 & $515.95\pm3.41$ & 1.38 & 7.05  & $140\pm13$  & $194\pm17$  &  23.54 & 37.26 & $14.02\pm1.30\pm1.08$ & $12.29\pm1.08\pm0.93$  &$13.07\pm0.83\pm0.70$ \\
  $4.410^{r}$  & $7.19\pm0.01$ & 1.14 & 7.05 & \multicolumn{2}{c}{$3.3^{+2.6}_{-1.8}$} & \multicolumn{2}{c}{31.75} & \multicolumn{3}{r}{ $10.00^{+8.39}_{-5.81}\pm0.82$}\\
  4.4156 & $1090.66\pm6.89$& 1.17 & 2.68 & $233\pm16$  & $401\pm24$  &  22.19 & 34.64 & $13.76\pm0.94\pm0.56$ & $15.18\pm0.91\pm0.57$  &$14.49\pm0.66\pm0.40$ \\
  $4.430^{r}$  & $6.96\pm0.01$   & 1.17 & 2.83 & \multicolumn{2}{c}{$4.0^{+2.2}_{-1.9}$} & \multicolumn{2}{c}{29.82} & \multicolumn{3}{r}{ $13.50^{+7.62}_{-6.58}\pm0.54$}\\
  4.4359 & $582.34\pm3.84$ & 1.16 & 2.83 & $110\pm12$  & $219\pm18$  &  22.23 & 35.27 & $12.17\pm1.33\pm0.51$ & $15.36\pm1.26\pm0.60$  &$13.81\pm0.92\pm0.39$ \\
  $4.450^{r}$  & $7.86\pm0.01$   & 1.19 & 2.83 & \multicolumn{2}{c}{$3.4^{+2.6}_{-1.9}$} & \multicolumn{2}{c}{29.32} & \multicolumn{3}{r}{ $10.42^{+7.97}_{-5.82}\pm0.41$}\\
  $4.460^{r}$  & $8.96\pm0.01$   & 1.19 &  10.22 & \multicolumn{2}{c}{$6.0^{+3.1}_{-2.4}$} & \multicolumn{2}{c}{29.14} & \multicolumn{3}{r}{ $16.11^{+8.32}_{-6.44}\pm1.70$}\\
  4.4671 & $111.09\pm0.73$ & 1.17 & 10.22 & $25\pm6$    & $26\pm7$    &  22.05 & 35.61 & $15.10\pm3.48\pm1.61$ & $9.38\pm2.52\pm0.99$   &$11.78\pm2.07\pm0.89$ \\
  $4.480^{r}$  & $8.39\pm0.01$   & 1.20 & 10.22 & \multicolumn{2}{c}{$2.6^{+2.9}_{-1.6}$} & \multicolumn{2}{c}{28.97} & \multicolumn{3}{r}{ $10.30^{+8.30}_{-4.58}\pm0.79$}\\
  $4.500^{r}$  & $8.24\pm0.01$   & 1.22 & 10.22 & \multicolumn{2}{c}{$0.4^{+1.7}_{-0.9}$} & \multicolumn{2}{c}{28.40} & \multicolumn{3}{r}{ $3.82^{+4.99}_{-2.64}\pm0.12$}\\
  $4.520^{r}$  & $8.95\pm0.01$   & 1.22 & 10.22 & \multicolumn{2}{c}{$2.8^{+2.2}_{-1.5}$} & \multicolumn{2}{c}{28.37} & \multicolumn{3}{r}{ $7.82^{+5.93}_{-4.04}\pm0.80$}\\
  4.5271 & $112.12\pm0.74$ & 1.12 & 2.84 & $25\pm6$    & $27\pm6$    &  23.59 & 38.18 & $14.08\pm3.38\pm0.59$ & $9.76\pm2.44\pm0.38$   &$11.57\pm2.00\pm0.33$ \\
  $4.540^{r}$  & $9.66\pm0.01$   & 1.23 & 2.84 & \multicolumn{2}{c}{$1.5^{+2.6}_{-1.9}$} & \multicolumn{2}{c}{28.20} & \multicolumn{3}{r}{ $11.98^{+6.49}_{-4.74}\pm0.15$}\\
  $4.550^{r}$  & $9.05\pm0.01$   & 1.23 & 2.84  & \multicolumn{2}{c}{$3.5^{+2.6}_{-1.9}$} & \multicolumn{2}{c}{28.13} & \multicolumn{3}{r}{ $11.74^{+6.93}_{-5.07}\pm0.36$}\\
  $4.560^{r}$  & $8.59\pm0.01$   & 1.24 & 2.84 & \multicolumn{2}{c}{$2.6^{+2.4}_{-1.7}$} & \multicolumn{2}{c}{28.01} & \multicolumn{3}{r}{ $8.99^{+6.74}_{-4.78}\pm0.28$}\\
  $4.570^{r}$  & $8.69\pm0.01$   & 1.24 & 2.84 & \multicolumn{2}{c}{$1.9^{+2.2}_{-1.5}$} & \multicolumn{2}{c}{28.71} & \multicolumn{3}{r}{ $5.93^{+5.93}_{-4.05}\pm0.20$}\\
  4.5745 & $48.93\pm0.32$  & 1.06 &  3.93  & $7\pm3$     & $21\pm5$    &  23.50 & 39.88 & $9.63\pm4.13\pm0.48$  & $16.21\pm4.05\pm0.77$  &$12.95\pm2.89\pm0.45$ \\
  $4.580^{r}$  & $8.84\pm0.01$   & 1.25 & 12.26 & \multicolumn{2}{c}{$3.9^{+2.8}_{-2.0}$} & \multicolumn{2}{c}{28.51} & \multicolumn{3}{r}{ $13.87^{+7.47}_{-5.34}\pm1.31$}\\
  $4.590^{r}$  & $8.50\pm0.01$   & 1.25 & 12.26 & \multicolumn{2}{c}{$0.9^{+1.8}_{-1.1}$} & \multicolumn{2}{c}{28.67} & \multicolumn{3}{r}{ $3.86^{+4.96}_{-3.03}\pm0.31$}\\
  4.5995 & $586.89\pm3.87$ & 1.49  &  12.26 & $64\pm9$    & $99\pm14$   &  19.37 & 31.79 & $6.35\pm0.89\pm0.80$  & $5.98\pm0.85\pm0.75$   &$6.16\pm0.61\pm0.55$  \\
  4.6119 & $103.83\pm0.68$ & 2.35  & 55.62 & $7\pm3$      & $5\pm4$       &  14.22 & 23.18 & $3.86\pm1.89\pm2.18$  & $1.80\pm1.50\pm1.00$   &$2.72\pm1.20\pm1.13$  \\
  4.6280 & $521.52\pm3.37$ & 1.22 & 7.34 & $43\pm8$    & $91\pm13$   &  18.55 & 31.20 & $6.09\pm1.13\pm0.49 $ & $7.68\pm1.10\pm0.61$   &$6.90\pm0.80\pm0.39$  \\
  4.6409 & $552.41\pm3.57$ & 1.43 & 8.72 & $54\pm8$    & $75\pm12$   &  19.77 & 32.22 & $5.79\pm0.86\pm0.54 $ & $4.95\pm0.79\pm0.46$   &$5.35\pm0.59\pm0.35$  \\
  4.6612 & $529.63\pm3.45$ & 1.48 & 3.75 & $42\pm7$    & $69\pm12$   &  18.87 & 31.09 & $4.73\pm0.79\pm0.23 $ & $4.74\pm0.82\pm0.22$   &$4.73\pm0.58\pm0.16$  \\
  4.6819 &$1669.31\pm10.77$& 1.35 & 2.59 & $134\pm14$  & $210\pm20$  &  18.41 & 30.66 & $5.40\pm0.56\pm0.22 $ & $5.10\pm0.49\pm0.19$   &$5.24\pm0.38\pm0.15$  \\
  4.6988 & $536.20\pm3.47$ & 1.33 & 3.97 & $46\pm8$    & $72\pm12$   &  18.65 & 31.97 & $5.77\pm1.00\pm0.29$  & $5.22\pm0.88\pm0.25$   &$5.48\pm0.68\pm0.19$  \\
\hline
\hline
\end{tabular}}
\label{totCrosssection}
\end{table*}

\end{document}